\documentclass[camera,letterpaper,nomarginnotes,nonarrowgutter,hyperref]{jpaper}

\usepackage{amsmath,amssymb,amsfonts}
\usepackage{algorithm}
\usepackage[noend]{algpseudocode}
\usepackage{algorithmicx}
\usepackage{graphicx}
\usepackage{textcomp}
\usepackage{xcolor}

\usepackage{makecell}
\usepackage{array}
\usepackage{booktabs}
\usepackage[sort,compress]{cite}

\usepackage{amsmath, amssymb, amsfonts}
\usepackage{graphicx}
\usepackage{textcomp}
\usepackage{url}
\usepackage{algorithm}
\usepackage{algpseudocode}
\usepackage{multirow}
\usepackage{tikz}
\usepackage{marginnote}
\usepackage{fancyhdr}
\usepackage{titlesec}
\usepackage{dblfloatfix}
\usepackage{makecell}
\usepackage{comment}
\usepackage{xspace}
\usepackage{xcolor, soul}
\usepackage{enumitem}
\usepackage{tabularx} 
\PassOptionsToPackage{bookmarks=true,breaklinks=true,letterpaper=true,colorlinks,citecolor=blue,linkcolor=blue,urlcolor=blue}{hyperref}
\usepackage{hyperref}
\usepackage{setspace}
\usepackage[colorinlistoftodos,prependcaption,textsize=scriptsize]{todonotes} 
\usepackage[sort,compress]{cite}

\usepackage{siunitx}
\usepackage{datetime}

\usepackage[dvipsnames]{xcolor}
\usepackage{soul}
\usepackage{paracol}
\usepackage[bottom]{footmisc}
\usepackage[most]{tcolorbox}
\usepackage{amsmath}

\usepackage{fancyhdr}
\usepackage{mathtools}

\usepackage{booktabs}

\usepackage{soul}

\definecolor{bleudefrance}{rgb}{0.19, 0.55, 0.91}
\definecolor{caribbeangreen}{rgb}{0.0, 0.8, 0.6}
\definecolor{english}{rgb}{0.0, 0.5, 0.0}
\definecolor{teal}{rgb}{0.0, 0.5, 0.5}
\definecolor{tyrianpurple}{rgb}{0.4, 0.01, 0.24}
\definecolor{violet}{rgb}{0.56, 0.0, 1.0}
\definecolor{darkorange}{rgb}{0.78, 0.43, 0}

\newcommand*\circled[1]{\tikz[baseline=(char.base)]{
            \node[shape=circle,draw,inner sep=0.4pt,fill=black, text=white] (char) {#1};}}

\newcommand{\squishlist}{
 \begin{list}{$\circ$}
  { \setlength{\itemsep}{0pt}
     \setlength{\parsep}{0pt}
     \setlength{\topsep}{3pt}
     \setlength{\partopsep}{0pt}
     \setlength{\leftmargin}{1em}
     \setlength{\labelwidth}{1em}
     \setlength{\labelsep}{0.5em} } }

\newcommand{\squishend}{  \end{list}  }

\definecolor{darkspringgreen}{rgb}{0.09, 0.65, 0.27}
\definecolor{darkspringgreen2}{rgb}{0.09, 0.45, 0.27}
\definecolor{denim}{rgb}{0.08, 0.38, 0.74}
\definecolor{darkolivegreen}{rgb}{0.33, 0.42, 0.18}
\definecolor{tangerine}{rgb}{0.95, 0.52, 0.0}
\definecolor{mahogany}{rgb}{0.75, 0.25, 0.0}
\definecolor{coolblack}{rgb}{0.0, 0.18, 0.39}
\definecolor{darkpink}{rgb}{0.88, 0.28, 0.54}

\definecolor{seagreen}{rgb}{0.18, 0.55, 0.34}

\definecolor{pred}{rgb}{0.7843, 0.0039, 0.3137} 

\definecolor{darkpink}{rgb}{0.88, 0.28, 0.54}
\definecolor{forestgreen}{rgb}{0.0, 0.27, 0.13}
\definecolor{amber}{rgb}{1.0, 0.49, 0.0}

\usepackage[compact]{titlesec}
\titlespacing\section{0pt}{4pt plus 2pt minus 2pt}{4pt plus 2pt minus 2pt}
\titlespacing\subsection{0pt}{4pt plus 2pt minus 2pt}{4pt plus 2pt minus 2pt}
\titlespacing\subsubsection{0pt}{4pt plus 2pt minus 2pt}{4pt plus 2pt minus 2pt}
\makeatletter
\g@addto@macro{\normalsize}{%
  \setlength{\abovedisplayskip}{5pt plus 0.5pt minus 1pt}
  \setlength{\belowdisplayskip}{5pt plus 0.5pt minus 1pt}
  \setlength{\abovedisplayshortskip}{2pt}
  \setlength{\belowdisplayshortskip}{2pt}
  \setlength{\intextsep}{2pt plus 1pt minus 1pt}
  \setlength{\textfloatsep}{2pt plus 1pt minus 1pt}
  \setlength{\skip\footins}{2pt plus 1pt minus 1pt}}
\makeatother

 \usepackage{setspace}

\usepackage[colorinlistoftodos,prependcaption,textsize=scriptsize]{todonotes}
\usepackage{todonotes}

\usepackage{blindtext,graphicx}
\usepackage[absolute]{textpos}
\usepackage{datetime}

\definecolor{seagreen}{rgb}{0.18, 0.55, 0.34}
\definecolor{ballblue}{rgb}{0.13, 0.67, 0.8}

\definecolor{darkgreen}{rgb}{0.0, 0.44, 0.34}

\sloppypar

\definecolor{dollarbill}{rgb}{0.52, 0.73, 0.4}

\definecolor{indiagreen}{rgb}{0.07, 0.53, 0.03}

\newcommand\revhid[1]{\todo[linecolor=magenta,backgroundcolor=magenta!15,bordercolor=magenta]{\textcolor{black}{\textbf{#1}}}}

\renewcommand\revhid[1]{}

\newcommand\marklabel[1]{{\noindent\hyperref[rev:#1]{\markqg{#1}}}}

\usepackage{amsmath,amssymb,amsfonts}
\usepackage{graphicx}
\usepackage{textcomp}
\usepackage{xcolor}
\def\BibTeX{{\rm B\kern-.05em{\sc i\kern-.025em b}\kern-.08em
    T\kern-.1667em\lower.7ex\hbox{E}\kern-.125emX}}

\usepackage{hyperref}
\usepackage{cleveref}
\usepackage{xspace}
\usepackage{comment}
\usepackage{booktabs}
\usepackage{subfig}
\usepackage{tikz}
\usepackage{threeparttable}
\usepackage{tabularx}
\usepackage{multirow}
\usepackage[misc]{ifsym}

\newcommand{\codename}{GenPairX\xspace}
\newcommand{\codealg}{GenPair\xspace}
\newcommand{\mypar}[1]{\noindent\textbf{#1.}\xspace}
\newcommand*\circledw[1]{\tikz[baseline=(myanchor.base)] \node[circle,fill=white,inner sep=0.9pt,draw=black,line width=0.2mm] (myanchor) {\color{black}\bfseries\footnotesize #1};}

\begin{document}

\title{GenPairX: A Hardware-Algorithm Co-Designed Accelerator\\ for Paired-End Read Mapping}
\newcommand {\affilHUA}[0]{\textsuperscript{$\dagger$}}
\newcommand {\affilETH}[0]{\textsuperscript{$\ddagger$}}
\newcommand{\affilMAr}[0]{\textsuperscript{$\mathsection$}}
\newcommand{\affilKIN}[0]{\textsuperscript{$\mathparagraph$}}
\newcommand{\corr}[0]{\textsuperscript{\Letter}}
\newcommand\CR[1]{{\color{black} {}#1}}
\newcommand\CRvII[1]{{\color{black} {}#1}}
\newcommand\kona[1]{{\color{black} {}#1}}
\newcommand\crviii[1]{{\color{black} {}#1}}
\newcommand\crviv[1]{{\color{black} {}#1}}

\def\hpcacameraready{} 
\newcommand{\hpcapubid}{0000--0000/00\$00.00}

\newcommand\hpcaauthors{Julien Eudine\affilHUA~~ Chu Li\affilHUA~ Zhuo Cheng\affilHUA~~ Renzo Andri\affilHUA~~  Can Firtina\affilETH\\ Mohammad Sadrosadati\affilETH~~ Nika Mansouri Ghiasi\affilETH~~ Konstantina Koliogeorgi\affilETH~~ Anirban Nag\affilHUA\\
Arash Tavakkol\affilHUA~~ Haiyu Mao\affilKIN~~ Onur Mutlu\affilETH~~ Shai Bergman\affilHUA~~ Ji Zhang\affilHUA\corr\\\\}
\newcommand\hpcaaffiliation{\affilHUA\emph{Huawei Technologies Switzerland AG}~~ \affilETH\emph{ETH Zurich}~~ \\ \affilKIN\emph{King’s College London}~~  \corr\emph{Corresponding Author}}



\author{
\hpcaauthors{}
\hpcaaffiliation{}
}

\fancypagestyle{camerareadyfirstpage}{%
  \fancyhead{}
  \renewcommand{\headrulewidth}{0pt}
  \fancyhead[C]{
    \ifdefined\aeopen
    \parbox[][12mm][t]{13.5cm}{}
    \else
      \ifdefined\aereviewed
      \parbox[][12mm][t]{13.5cm}{}
      \else
      \ifdefined\aereproduced
      \parbox[][12mm][t]{13.5cm}{}
      \else
      \parbox[][0mm][t]{13.5cm}{}    \fi 
    \fi 
    \fi 
    \ifdefined\aeopen 
      \includegraphics[width=12mm,height=12mm]{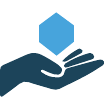}
    \fi 
    \ifdefined\aereviewed
      \includegraphics[width=12mm,height=12mm]{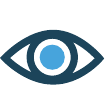}
    \fi 
    \ifdefined\aereproduced
      \includegraphics[width=12mm,height=12mm]{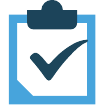}
    \fi
  }
  \fancyfoot[C]{}
}
\fancyhead{}
\renewcommand{\headrulewidth}{0pt}

\maketitle

\ifdefined\hpcacameraready 
  \thispagestyle{camerareadyfirstpage}
  \pagestyle{empty}
\else
  \thispagestyle{plain}
  \pagestyle{plain}
\fi

\newcommand{\hpcaheight}{0mm}
\ifdefined\eaopen
\renewcommand{\hpcaheight}{12mm}
\fi


\setcounter{page}{1}
\begin{abstract}

  Genome sequencing has become a central focus in computational biology due to its critical role in applications such as personalized medicine, disease outbreak tracking, and evolutionary research. A genome study typically begins with sequencing, which produces millions to billions of short DNA fragments known as reads. Extracting meaningful biological insights from these reads requires a computationally intensive step called read mapping, where each read is aligned to a reference genome. Read mapping \crviv{for short reads} comes in two forms: single-end and paired-end, with the latter being more prevalent due to its higher accuracy and support for advanced analysis. Read mapping remains a major performance bottleneck in genome analysis as a result of the extensive use of computationally intensive dynamic programming. Prior efforts have attempted to mitigate this cost by \crviv{employing 
filters to identify and potentially discard computationally expensive matches and leveraging hardware accelerators to speed up the computations.}
While partially effective, these approaches have limitations. In particular, existing filters are often ineffective for paired-end reads, as they evaluate each read independently and exhibit \CR{relatively low} filtering ratios.

In this work, we propose \codename{}, a hardware-algorithm co-designed accelerator that efficiently minimizes the computational load of paired-end read mapping while enhancing the throughput of memory-intensive operations. \codename{} introduces: (1) a novel filtering algorithm that jointly considers both reads in a pair 
to improve filtering effectiveness, and a lightweight alignment algorithm to replace \CR{most} of the computationally expensive dynamic programming operations, and (2) two specialized hardware mechanisms to support the proposed algorithms. The proposed hardware addresses the high memory bandwidth demands of the read filtering process via orchestration of memory accesses over high-bandwidth memory channels, and accelerates the alignment of candidate reads via simple vectorized logical XOR operators.
Our evaluations show that \codename{} delivers substantial performance improvements over state-of-the-art solutions, achieving 1575$\times$  and 1.43$\times$ higher throughput per watt compared to leading CPU-based and accelerator-based read mappers, respectively, all without compromising accuracy.

\end{abstract}

\section{Introduction}

Genome study is essential for \CR{personalized medicine~\cite{Kingsmore2024,cells13060504,clark2019diagnosis,farnaes2018rapid,sweeney2021rapid,alkan2009personalized,ginsburg2009genomic,chin2011cancer}, disease outbreak tracking~\cite{Bandoy2021, ZHENG2024121513,bloom2021massively,yelagandula2021multiplexed,le2013selected,nikolayevskyy2016whole,qiu2015whole,gilchrist2015whole},  agriculture~\cite{prasad2021soil,Mascher2024,Schreiber2024}, scientific discovery~\cite{urbanek2018degradation,edgar2022petabase,paoli2022biosynthetic}, and evolutionary studies~\cite{02233705-202312000-00003, doi:10.1128/spectrum.03617-22, Chiu2023,shi2018evolutionary}.}
\crviv{Modern genome studies involve two major stages: \textit{sequencing} and \textit{analysis}}. In the sequencing stage, \CR{raw DNA or RNA} is fragmented and read by high-throughput sequencing machines, producing billions of short sequences, namely \emph{reads}. The analysis stage processes this data to reconstruct the genome, identify genetic variants, and extract biologically meaningful insights, requiring intensive computation and memory accesses across multiple software pipelines. Among sequencing techniques, \emph{paired-end sequencing} has been widely adopted, used in over 90\% of sequencing workflows today\cite{werner2012comparison,bartram2011generation,campbell2008identification,borsting2015next,rizzo2012key,edgren2011identification}.\footnote{Sequencing technologies continue to evolve to meet the growing demands of genome analysis applications, with diverse platforms spanning high-accuracy short-read approaches from Illumina\cite{illumina} to long-read technologies offered by Pacific Biosciences\cite{PacBio2025} and Oxford Nanopore Technologies\cite{NanoporeTech2025}, each addressing different analytical needs in resolution, scale, and structural insight. This work focuses on \emph{short} read mapping, and particularly optimizing read mapping for \emph{short paired-end reads}.} In paired-end sequencing, both ends of each DNA fragment are read, generating two sequences per fragment with a known orientation and approximate distance between them. This provides increased context for each read pair, enabling higher alignment accuracy, improved resolution of structural variants, and better handling of repetitive or \CR{ambiguous genomic regions~\cite{illumina_paired_end_2025, https://doi.org/10.1002/0471250953.bi1506s45, Freedman2020, Zhu2015, 10.1093/bioinformatics/btq152,eren2013filtering,grimm2013accurate,cameron2019comprehensive,rausch2012delly}}.

A central and computationally demanding task in the analysis stage is read mapping, where reads are aligned to a reference genome. This involves a mix of computationally intensive tasks like string matching, indexing, and dynamic programming (DP), making it a major bottleneck and a prime target for \CR{acceleration~\cite{mutlu2023accelerating,alser2022molecules,alser2020accelerating}}. 
To address these bottlenecks, prior work has explored a variety of hardware techniques that can be categorized into two main groups. First, several works propose less compute-intensive, \CR{coarse-grained filters~\cite{10.1145/3503222.3507702, alser2020sneakysnake, alser2017gatekeeper,xin2013accelerating,xin2015shifted,kim2018grim,alser2019shouji}} for single-end reads (i.e., non-paired-end reads) to quickly filter out dissimilar sequences. \crviv{These filters effectively reduce the workload of fine-grained compute-intensive operations (e.g., DP), thereby increasing the performance of read mapping.
Second, many works design specialized hardware-based accelerators, such as FPGA-based~\cite{Liyanage2023,6239809,Samarasinghe2021Energy,alser2019shouji,alser2020sneakysnake,alser2017gatekeeper,koliogeorgi2022gandafl,guo2019hardware,singh2021fpga,chen2021high,fei2018fpgasw,haghi2021fpga,rucci2018swifold,li2021pipebsw,liao2018adaptively}, ASIC-based~\cite{fujiki2018genax, nag2019gencache,turakhia2018darwin,Lindegger2023Scrooge,cali2020genasm,banerjee2018asap,fujiki2020seedex,cali2022segram} and GPU-based~\cite{liu2010cudasw++,liu2013cudasw++,nishimura2017accelerating} accelerators, to directly accelerate the performance bottlenecks.}

Unfortunately, existing state-of-the-art filters and DP accelerators are \emph{not} \crviv{optimized} for \emph{paired-end read mapping}, \CR{a major} step in paired-end \CR{sequence analysis}. This step presents significant computational challenges, as it involves aligning \CR{\emph{both}} ends of a DNA fragment to a reference genome while maintaining their relative orientation and expected insert size.  Based on our experiments (detailed in \CR{\S\ref{sec:analysis}}), we identify two key reasons.
First, the portion of the reads that can be filtered is directly impacted if the dataset is paired-end reads. For example, the filtering efficiency of a state-of-the-art exact matching filtering technique drops from 55.7\% to 36.8\% when using paired-end reads instead of single-end reads \CR{(\S\ref{subsec:analysis:dist})}. Second, applying filters shifts read mapping from being compute-intensive to being memory-intensive \CR{(\S\ref{sec:seedmap})}, as filters are memory-intensive due to the large number of random memory accesses they require.

\textbf{The goal} of this work is to improve the performance of the widely used \CR{\emph{paired-end read mapping}} algorithm by (1) increasing the effectiveness of the read filtering stage to discard a \CR{large} fraction of redundant computations, and (2) accelerating \CR{overall} execution by leveraging techniques with lower computational complexity than traditional DP approaches.  To this end, we propose \textit{\codename}, the first paired-end read mapping accelerator via hardware-algorithm co-design. The \textbf{key idea} behind \codename is to leverage the spatial locality of exact matches and small variations in paired-end reads, while minimizing redundant computations and memory accesses.

\codename{} introduces innovations at both the \CR{algorithm} and hardware levels. (1) \CR{\underline{Algorithm level}}: We present \codealg, a novel paired-end read mapping algorithm that includes a new hash-based filter specifically designed for paired-end reads, along with a lightweight alignment method. \codealg's mapping is motivated by the observation that both reads in the pair are mapped to the same region of the reference genome, with both reads containing partial exact matches to the reference. Consequently, the hash-based filter queries an index of mapping locations with partial exact matches for each read. If some locations from both reads fall within the same region of the reference genome, they are retained as potential mapping candidates.
\codealg's lightweight alignment algorithm is based on the observation that 69.9\% of read-pairs exhibit simple edits that consist solely of mismatches or consecutive insertions or deletions.
Hence, \codealg by default performs simple XOR operations, replacing costly DP operations with computationally simpler alternatives to reduce alignment time. (2) \CR{\underline{Hardware level}}: We introduce \codename{}, a hardware architecture tailored to support and accelerate \codealg operations. \codename{} incorporates two key mechanisms. First, it accelerates memory accesses for the hash-based filter by leveraging the maximum available bandwidth per memory channel. Second, it accelerates the lightweight XOR alignment algorithm by vectorizing the underlying logic operations.

We carefully design a high-throughput, balanced end-to-end read mapping pipeline and evaluate the impact of the proposed algorithmic and hardware contributions on the
end-to-end performance of paired-end read mapping. 
We integrate \codename{} with GenDP\CR{~\cite{gu2023gendp}}, a hardware accelerator that serves as a fallback mechanism for the small fraction of read-pairs that cannot be mapped or aligned by \codename{}.
\codename{}\CR{'s} filtering mechanism and lightweight alignment approach successfully maps 89.1\% and aligns 76.1\% of the reads without relying on \CR{computationally-intensive} dynamic programming.
We compare the proposed pipeline against the state-of-the-art software Minimap2~\cite{li2018minimap2} (CPU implementation), an end-to-end GPU implementation of \CR{the BWA-MEM\cite{houtgast2017efficient} read mapping pipeline,} and the state-of-the-art ASIC-based accelerator, GenCache~\cite{nag2019gencache} (more details and systems are discussed in \CR{\S\ref{sec:methodology})}.
Our experimental results show that \codename{} offers substantial improvements in throughput, energy efficiency, and area efficiency, achieving end-to-end \CR{throughput per Watt improvements} of 1575$\times$ and 1.43$\times$ compared to \CR{state-of-the-art} CPU and ASIC read mappers, respectively. \codename{} improves throughput per area by 958$\times$ and 2.38$\times$ compared to \CR{state-of-the-art} CPU and ASIC mappers while maintaining or \CR{improving} mapping accuracy.

\noindent This paper makes the following contributions:
\begin{itemize}
    \item We perform an extensive profiling of paired-end read mapping and identify limitations and inefficiencies of commonly-used mapping tools when handling \CR{\emph{paired-end}} reads.
    \item We introduce GenPair, a new read mapping algorithm for paired-end reads. GenPair is based on a novel hash-based filter tailored for paired-end reads and a lightweight alignment algorithm that leverages the patterns in the edit variations of the expected alignments to replace unnecessary expensive dynamic programming operations with more efficient ones.
    \item We introduce \codename{}, the first hardware-algorithm co-designed accelerator for \CR{\emph{paired-end}} read mapping. \codename{} addresses the memory bottleneck inherent in \codealg by designing an optimized data structure, as well as an efficient hardware pipeline to achieve high performance and power efficiency.
    \item We evaluate \codename{} against \CR{two} state-of-the-art software and hardware read mappers, demonstrating large \textit{performance per Watt} and\textit{ performance per area} improvements while maintaining or improving accuracy.
\end{itemize}

\section{Background}
\label{sec:background}

\mypar{Paired-end read mapping}\label{sec:back_read}
Paired-end reads are DNA sequences obtained from opposite ends of the same DNA fragment.
Unlike single-end read mapping, paired-end mapping involves aligning \CR{\emph{both}} reads of a pair to the reference genome, taking into account the distance of their mapped locations in the reference genome.
This distance is expected to be small, providing an additional constraint for alignment~\cite{li2018minimap2}.

\mypar{Paired-end read mapping process}
Paired-end read mapping consists of three steps: seeding, chaining, and sequence alignment.

\noindent\emph{Seeding.} 
In the seeding step, the read mapper extracts \emph{seeds}, i.e., fractions of a read, and queries the reference genome's helper index data structure to locate their occurrences in the reference genome. Matches in the index are recorded as candidate mapping positions, which are then further evaluated in subsequent steps.

\noindent\emph{Chaining.} 
In the chaining step, the seeds are filtered and merged into longer regions called chains. This process uses \CR{dynamic programming (DP)\cite{baichoo2017computational}} to calculate the chaining score, which quantifies the degree of alignment between the read and the reference genome. The goal is to chain seeds to ensure that they form a coherent alignment. 

\noindent\emph{Sequence alignment.} 
The sequence alignment step uses the chains to compute the optimal alignment of a read-pair to the reference genome, accounting for mismatches, insertions, and deletions, typically using DP-based algorithms like Smith-Waterman~\cite{SMITH1981195} \CR{or Needleman-Wunsch\cite{needleman1970general}}.
\crviii{The final alignment \crviv{is} encoded using CIGAR strings (Compact Idiosyncratic Gapped Alignment Report)\cite{Li2009SAM}, which is a compressed code in SAM/BAM\cite{Li2009SAM} files that describes how a short DNA sequence (a read) aligns to a longer reference genome, using letters and numbers to show matches, mismatches, insertions, and deletions.}

Chaining and sequence alignment are the most computationally intensive steps of the \CR{read mapping process} due to their reliance on dynamic programming. Paired-end read mapping further amplifies this complexity due to the increased number of candidate alignments that must be evaluated, i.e., the distance between reads in a pair must be considered to form valid chains, while sequence alignment must account for \CR{\emph{both}} reads' relative positions.
For example, prior literature observes that for paired-end reads, the chaining stage consumes over 65\% of the total execution time, while the alignment stage accounts for more than 16\%~\cite{Alser2025}.

\section{Motivation and Insights: Analysis of Paired-End Read Mapping}
\label{sec:analysis}

Paired-end read mapping is a technique in genome analysis where both ends of DNA fragments are sequenced and aligned to a reference genome. It surpasses the mapping accuracy of single-end reads, as it is found to enable more accurate detection of structural variants, insertions/deletions, and repetitive regions~\cite{illumina_paired_end_2025, https://doi.org/10.1002/0471250953.bi1506s45, Freedman2020, Zhu2015, 10.1093/bioinformatics/btq152,eren2013filtering,grimm2013accurate,cameron2019comprehensive,rausch2012delly}.
For example, results from performing the same read mapping task with single-end and paired-end reads (see \S\ref{sec:methodology} for details 
of the tools and datasets used in our experiments), show that paired-end read mapping yields substantially higher accuracy. Specifically, the $F_1$ score for SNPs declines from 0.9913 to 0.9820 when switching from paired-end to single-end mapping, and for insertions and deletions (INDELs), it drops from 0.9326 to 0.9090.
These advantages have driven the increasing adoption of short-read sequencing in genomics research~\cite{illumina_paired_end_2025}, and have led major manufacturers to focus on producing paired-end read sequencing platforms~\cite{mordorintelligence_short_read_sequencing_2025,iontorrent}.

In this section, we first analyze the performance bottlenecks of paired-end read mapping (\S\ref{subsec:analysis:bottleneck}) and then present three key observations (\S\ref{subsec:analysis:dist}-\S\ref{sec:obs3}) that reveal potential optimization opportunities in paired-end mapping.
Our profiling methodology is based on Minimap2~\cite{li2018minimap2}, a state-of-the-art tool for read mapping. We leverage three commonly-used paired-end read datasets \CR{(see \S\ref{sec:methodology})}, each comprising 150 base pairs (bp), sourced from the Genome in a Bottle (GIAB) project~\cite{giab_hg002, zook2016extensive}, using GRCh38~\cite{giab_grch38} as the human reference genome. 
\CR{Performance} bottlenecks and observed opportunities in paired-end mapping are as follows.

\begin{figure}[t]
\centering
\includegraphics[width=1.0\columnwidth]{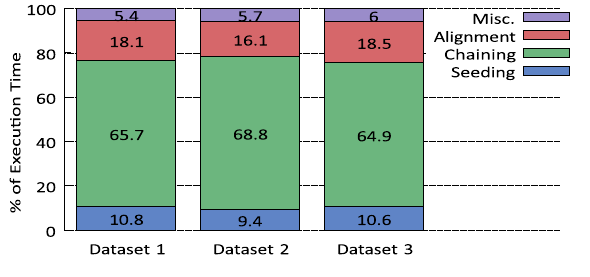}
\vspace{-1.5em}
\caption{Execution time breakdown (\%) of the state-of-the-art read mapper, Minimap2, when mapping paired-end reads.}
\label{fig:breakdown}
\end{figure}

\subsection{Performance Bottlenecks}\label{subsec:analysis:bottleneck}

We measure the time spent by Minimap2 in each stage of the paired-end read mapping process and present the results in \CR{Fig.}~\ref{fig:breakdown} for all three datasets. For each dataset, the two most compute-intensive stages, chaining and alignment, both of which rely on dynamic programming, account for between 83.4\% and 84.9\% of the total execution time. We therefore conclude that DP-based chaining and alignment are the performance bottleneck in the paired-end read mapping.

\subsection{Exact Match Distribution - Opportunity 1}\label{subsec:analysis:dist}
Prior work on single-end read filters capitalizes on the high accuracy of short reads~\cite{alser2022molecules} by looking for \emph{exact matches} of the \textit{entire} short reads against the reference genome~\cite{10.1145/3503222.3507702, nag2019gencache}, to avoid the computationally intensive process of chaining and alignment~\cite{9154510,mutlu2023accelerating}. For example,  we observe an exact match rate of 55.7\% for single-end reads in the human genome datasets. 
However, our profiling shows that the exact match rate drops to 36.8\% in the case of paired-end reads, as both reads in a pair must individually match the reference genome. As a result, a significant proportion of reads still require the computationally expensive steps of chaining and alignment.

We investigate how the high accuracy of short reads can be exploited to reduce dependence on chaining operations and DP-based alignments. To this end, we identify exact matches within segments of the paired-end reads and determine an optimal seed length that maximizes the exact match rate. This enables the efficient construction of longer contiguous sequences, facilitating more effective alignment across extended genomic regions.

\CRvII{\emph{Observation 1: At least one non-overlapping 50\,bp segment (i.e., either the first, middle, or last segment) is completely identical to the reference in both reads in 86.2\%, 85.8\%, and 84.9\% of cases for the three datasets respectively.} }

\emph{Insight:} This observation highlights the benefit of employing \CRvII{a seed matching mechanism that leverages long seeds for paired-end reads: it leads} to an efficient read mapping algorithm that accurately finds the mapping locations of a substantial proportion of read-pairs while avoiding compute-heavy chaining. This algorithm is described in details in \S\ref{sec:genpair}.

\subsection{Number of Seed Mapping Locations  - Opportunity 2}\label{sec:obs2}
\CRvII{The use of 50\,bp non-overlapping seeds (as described in \S\ref{subsec:analysis:dist}) leads to a considerable number of potential mapping locations to the reference genome. Experiments on our three datasets help us quantify this number.}

\CRvII{\emph{Observation 2: Across the three datasets, the average number of matching locations per seed is 9.6, 9.5, and 9.3, respectively for each dataset.}}

\emph{Insight:}  \CRvII{The consistently high number of mapping locations suggests challenges and inefficiency during the seed lookup queries.} To efficiently map the seeds, a specialized data structure is required to associate each seed with its corresponding potential mapping locations. The number of locations stored within this data structure directly impacts its design and implementation. We present this structure in \S\ref{sec:seedmap}.

\subsection{Edit Patterns in High-Scoring Read-Pairs  - Opportunity 3} 
\label{sec:obs3}
\begin{figure}[h]
\centering
\includegraphics[width=1.0\columnwidth]{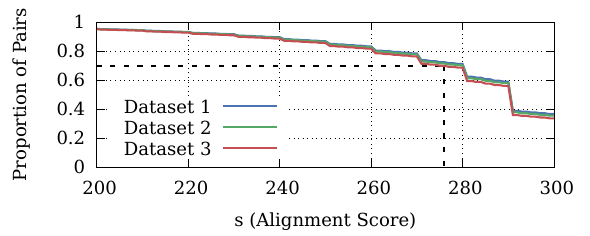}
\vspace{0em}
\caption{CDF of the min. align. score of both reads in a pair.}
\label{fig:observation1}
\end{figure}

We analyze the types of edits found in read-pairs with high alignment score, which quantifies the similarity between the read and the reference genome. To ensure consistency, we adopt the same scoring scheme as Minimap2 for short-read alignment, using affine gap penalties. A perfect match corresponds to a score of 300. Based on the scoring scheme, each edit type is penalized and leads to a lower score.
We use an alignment score of 276 as a threshold for high quality alignments, which excludes variations with two different types of edits. 
We enumerate all possible types of variations in a read sequence that result in a score equal or higher to the threshold. We present these variations 
in Table \ref{tab:score}, while
Fig. \ref{fig:observation1} illustrates the distribution of alignment scores for the read pairs for the three datasets.

\begin{table}[h]
    \centering
    \begin{tabular}{c|c}
        \toprule
        \textbf{Edit(s)} & \textbf{Alignment Score} \\
        \midrule
        None & 300 \\
         1 Mismatch & 290  \\
         1 Deletion & 286 \\
         1 Insertion & 284 \\
         2 Consecutive Deletions & 284 \\
         3 Consecutive Deletions & 282 \\
         2 Mismatches & 280 \\
         2 Consecutive Insertions & 280 \\
         4 Consecutive Deletions & 280 \\
         5 Consecutive Deletions & 278 \\
         1 Mismatch \& 1 Deletion & 276 \\
         \bottomrule
    \end{tabular}
    \caption{All edits that can occur in a read for alignment scores above 276. All reads with an alignment score strictly greater than 276 exhibit only a single type of edit.}
    \label{tab:score}
\end{table}

\emph{Observation 3: Across the three  datasets, 69.9\% of read-pairs exhibit edits that consist solely of mismatches or consecutive insertions or deletions, representing a limited subset of potential edit variations.}

\emph{Insight:} This observation enables us to develop a lightweight alignment algorithm that efficiently aligns reads with a single type of edit
without the need to rely on traditional dynamic programming methods. We present this algorithm in detail in \S\ref{sec:light_align}.
\section{\CRvII{\codealg{} Algorithm}}
\noindent\crviii{We present \codealg, a novel hardware-optimized algorithm, based on our key observations and insights outlined in \S\ref{sec:analysis}}.

\subsection{Overview} \label{sec:genpair}
\begin{figure}[h]
\centering
\includegraphics[width=1.0\columnwidth]{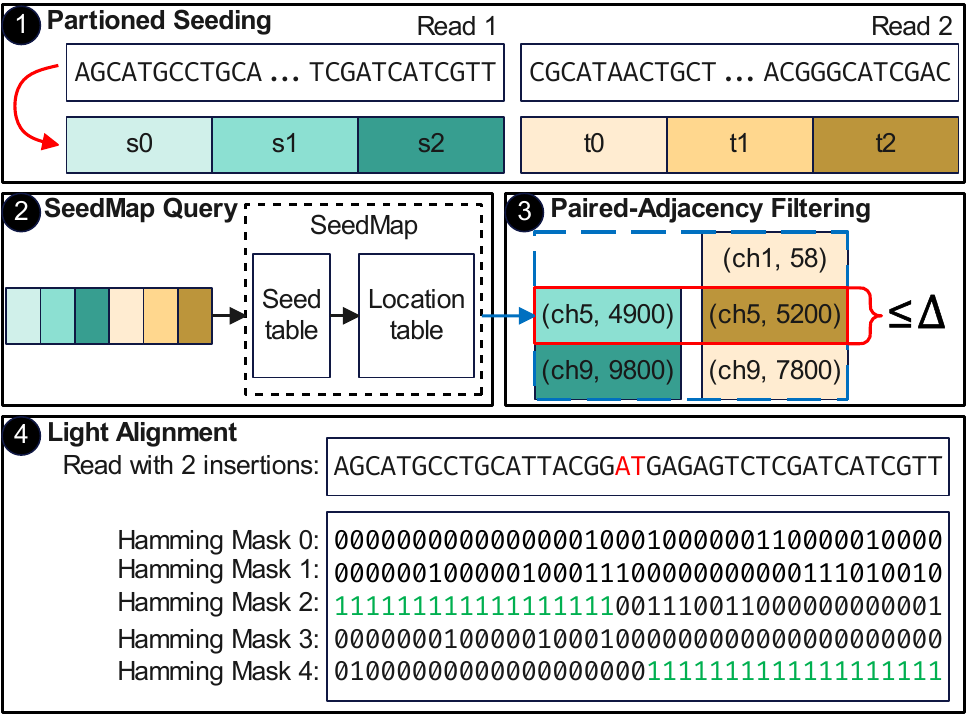}
\caption{\crviii{Overview of the four online steps of the \codealg algorithm for read mapping.}}
\label{fig:votemap}
\end{figure}

\crviii{\codealg operates in two stages: offline and online. \textit{Offline Stage:} This stage corresponds to the indexing of the reference genome. We construct \textit{SeedMap}, a compact hash-table structure that serves as an index table for efficient lookups to the reference genome (\S\ref{sec:seedmap}).   
\textit{Online Stage:} This stage implements the mapping of paired-end reads to the reference genome, leveraging SeedMap.
Fig.~\ref{fig:votemap} provides a high level overview of the \crviv{online \codealg read mapping pipeline}, which is composed of four steps:
First, the \emph{Partitioned Seeding} step~\circled{1} extracts seeds from the read-pair and encodes them into hash values (\S\ref{sec:algo:partitioned_seeding}). The subsequent \emph{SeedMap Query} step~\circled{2} uses these seeds to query SeedMap and identify potential matching locations in the reference genome (\S\ref{sec:algo:seedmapquery}).
The retrieved locations are then passed to the \emph{Paired-Adjacency Filtering} step~\circled{3}, which filters out mapping locations that are unlikely to deliver a valid alignment (\S\ref{sec:algo:filtering}). This step also fulfills the traditional chaining role, by extending the remaining unfiltered seed hits into the sequences to be aligned. Finally, the \emph{Light Alignment} step~\circled{4} aligns each read in the pair to the reference genome at the candidate locations identified in the previous stage (\S\ref{sec:light_align}).
We describe the offline SeedMap construction and each step of the online read mapping pipeline in the following subsections.}

\begin{figure*}[b]
  \centering
  \subfloat[SeedMap construction and data structure.\label{fig:index_building}]{
    \includegraphics[width=0.65\textwidth]{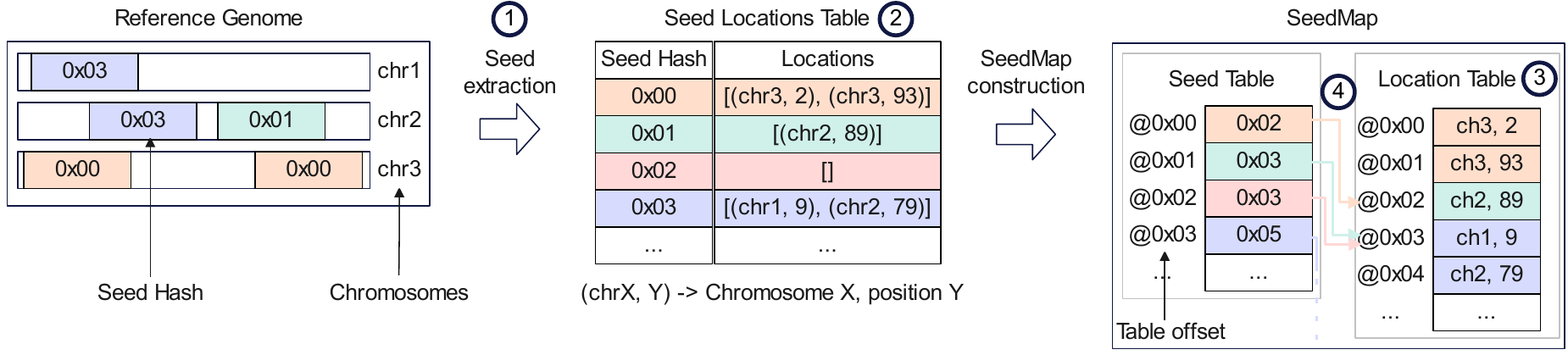}
  }\hfill
  \subfloat[\crviv{Online query} process of SeedMap.\label{fig:hashmap}]{
    \includegraphics[width=0.32\textwidth]{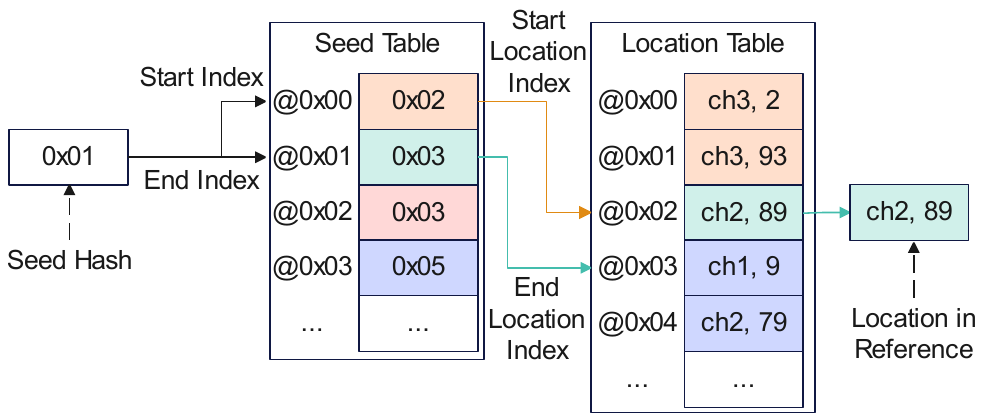}
  }
  \caption{\crviv{Overview of offline SeedMap construction and online SeedMap querying}.}
  \label{fig:seedmap_overview}
\end{figure*}

\subsection{Offline SeedMap Construction}\label{sec:seedmap}
The construction of SeedMap takes place offline, i.e., before the mapping. It is done only once for a given reference genome, and SeedMap can be reused when mapping different read datasets against the same reference genome.

\noindent\textit{SeedMap Design and Construction.} \CRvII{SeedMap resembles a hash-table index structure and comprises two tables: the Location Table and the Seed Table. The \textbf{Location Table} stores all seed locations grouped by seed.
The \textbf{Seed Table} serves as an index \crviv{into} the Location Table. Each entry in the Seed Table corresponds to a specific seed hash value and points to the offset in the Location Table that includes the first location of this seed in the reference genome.}
\CR{Fig.}~\ref{fig:index_building} illustrates the SeedMap data structure and its construction using an example.
The construction begins with extracting seeds from the reference genome and assigning each seed a hash value~\circledw{1}. These seeds are sorted by their hash values and organized into the \textit{temporary} Seed Locations Table~\circledw{2}, which includes all seed locations in the reference genome, i.e., the chromosome and the offset within the chromosome. Based on this data, we then build the Location Table~\circledw{3} by linearly concatenating all the locations. Thanks to the sorted hash values, the reference genome locations associated with the same seed are stored contiguously in memory. We then compute the position of the first location for each seed in the Location Table and build the Seed Table~\circledw{4} accordingly.

Our data structure builds upon prior works such as Darwin~\cite{turakhia2018darwin}, which uses two chained associative arrays. While we adopt the same overall structure, our implementation is further tuned to improve memory efficiency and minimize memory accesses. \crviii{We achieve this with two key techniques. First, we index the Seed Table using a \textit{hash of the seed} rather than storing the full seed sequence, to reduce the memory footprint. Second, we adopt a data layout that stores the reference genome locations associated with the same seed \textit{contiguously}, allowing us to minimize memory accesses and enable efficient burst transfers.}

\subsection{Partitioned Seeding}
\label{sec:algo:partitioned_seeding}
The objective of the Partitioned Seeding stage is to extract seeds from each read-pair and encode the DNA segment into a hash value.  
As illustrated in~\circled{1} of \CR{Fig.}~\ref{fig:votemap}, the Partitioned Seeding mechanism extracts six seeds per read-pair. Our seeding strategy extracts 50 bases-long non-overlapping seeds, i.e., three seeds per read corresponding to the first, middle, and last 50\,bp of each read. Based on Observation 1 of \S\ref{subsec:analysis:dist}, in 86\% of the cases, at least one seed from each pair matches exactly the reference genome, \crviii{enabling SeedMap to accurately return} the true locations for the reads. Each seed is encoded into a 32\,bit value by applying the xxHash~\cite{xxhash2024} function.

\subsection{\CRvII{SeedMap Query}}
\label{sec:algo:seedmapquery}

The SeedMap Query stage (\circled{2} in \CR{Fig.}~\ref{fig:votemap}) operates on the hashed seeds generated in the previous step and retrieves \crviii{their} potential locations in the reference genome. 
The querying process for SeedMap, illustrated in \CR{Fig.}~\ref{fig:hashmap}, begins by using the seed hash value as an index to access the Seed Table.
The current and \crviii{previous} index of the Seed Table correspond to a continuous region in the Location Table, that includes all possible locations in the reference genome where the seed can be mapped. Note that the possible locations for each seed are laid out sequentially in memory and already sorted during the construction of SeedMap. Consequently, when querying SeedMap with the three seeds from a read in a read-pair, we only need to merge the three sorted location lists. This design maximizes data locality through consecutive memory accesses to the Location Table for a single seed, optimizing query throughput.

\subsection{Paired-Adjacency Filtering}
\label{sec:algo:filtering}
The goal of the Paired-Adjacency Filtering  (\circled{3} in \CR{Fig.}~\ref{fig:votemap}) is to identify the potential matching locations that lead to a high likelihood of a correct mapping, avoiding redundant computations.
This approach is inspired by the adjacency filtering method of \crviii{FastHASH~\cite{xin2013accelerating, fasthashsource}}, where consecutive seeds within a single read are checked to see if they map to adjacent positions in the reference genome.

Paired-end reads demonstrate an inherent adjacency, i.e., a typical positional distance between their reference genome locations of less than a dataset-defined threshold, $\Delta$, usually 200 to 500 bp. Our Paired-Adjacency Filtering \crviii{technique} leverages this feature and only considers candidates whose distance lies within this $\Delta$ threshold.
Paired-Adjacency Filtering algorithm operates in two steps: (1) it fetches the sorted seed location lists derived from read 1 and read 2 during SeedMap Query,  and (2) iterates simultaneously over both lists to compare their locations. When the distance between a location from read 1 and a location from read 2 is below $\Delta$, the location is added to the list of potential mapping sites.
The output of the Paired-Adjacency filtering stage provides potential alignment positions in the reference genome, each of them corresponding to a perfect match with a 50 bp seed segment from the read.

\subsection{Light Alignment}\label{sec:light_align}

The goal of the Light Alignment \crviii{step} (\circled{4} in \CR{Fig.}~\ref{fig:votemap}) is to efficiently align sequences with a limited number of edits, without relying on DP based methods. To achieve this, we leverage our Observation 3, which shows that  69.9~\% of all read pairs exhibit \crviv{one of the edit types} listed in \CR{Table}~\ref{tab:score}. The key idea is to examine whether the aligned sequences match any of the cases listed in \CR{Table}~\ref{tab:score}, starting with the one with the best score, until we find a valid one. If none is found, we resort to DP alignment.

Light Alignment is inspired by the \crviii{Shifted Hamming Distance (SHD) technique~\cite{xin2015shifted,shdsource}}, a filtering technique that has been broadly adopted by many prior works\crviv{, (e.g.,~\cite{alser2020sneakysnake, alser2017gatekeeper, alser2019shouji})} and uses parallel bitwise operations to eliminate dissimilar sequences and reduce alignment costs. Its core technique involves computing the Hamming mask between the read and several shifted copies of the reference sequence, allowing it to filter sequences that exceed a user-defined error threshold. Our approach extends SHD’s functionality beyond the scope of filtering by computing both the alignment score and the CIGAR string\cite{Li2009SAM}, thereby enhancing its applicability to alignment tasks.

\begin{figure}[h]
\centering
\includegraphics[width=1.0\columnwidth]{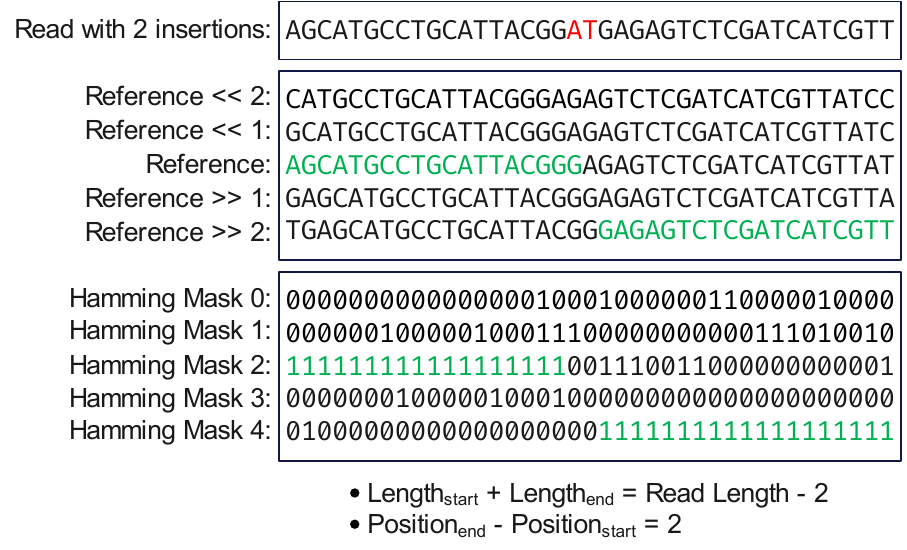}
\vspace{-1.5em}
\caption{Example of the Light Alignment mechanism.}
\label{fig:light_align}
\end{figure}

\CR{Fig.}~\ref{fig:light_align} illustrates the Light Alignment mechanism using an example read with two insertions (highlighted in red). It follows a three-step process: (1) Compute the Hamming mask between the read and several reference copies, each shifted by a specified number of base pairs. Each shift corresponds to a different edit variation. (2) Identify the Hamming mask with the longest consecutive segment of `\texttt{1}'s from both the start and end positions (highlighted in green). (3) Calculate the edit value by summing the lengths of these longest start and end segments of `\texttt{1}'s. The edit type is determined by the positional difference between these segments, and the position of the edit corresponds to the length of the longest start segment of `\texttt{1}'s.
In order to detect \textit{e} consecutive mismatches, insertions or deletions, we need at least $2e+1$ Hamming masks.
\vspace{0.5em}

\vspace{0.5em}
The Light Alignment method also supports identifying non-consecutive mismatches.
To detect this type of edit, we calculate the Hamming distance between the read and the reference by comparing each base-pair of the read to the reference.

\subsection{\CRvII{Support for Long-Read Data}}
\label{sec:longreads}

The long-read mapping problem can be reformulated as a paired-end mapping problem, enabling the direct application of the mapping component of \codealg. Each long read is partitioned into a sequence of interleaved read pairs, where the distance between the two reads in each pair is constrained to be less than $\Delta$. We then apply our \textit{Partitioned Seeding}, \textit{SeedMap Query}, and \textit{Paired-Adjacency Filtering} stages to each pair independently. For all pairs derived from the same long read, this process yields a set of candidate mapping locations on the reference genome. To further reduce false positives, we apply the Location Voting algorithm~\cite{Alser2025} across all candidate locations associated with the same read. Because long reads exhibit higher sequencing noise and a greater prevalence of structural variations than short reads~\cite{alser2022molecules}, light alignment is insufficient. Therefore, we perform full dynamic programming (DP) alignment on the resulting candidate regions.

\section{\codename Hardware Accelerator Design}
\label{sec:GA:HW}
\begin{figure}[h]
\centering
\includegraphics[width=1.0\columnwidth]{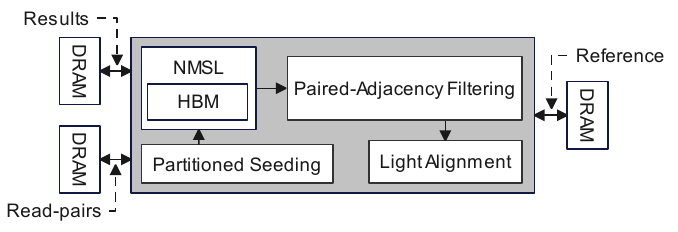}
\vspace{-1em}
\caption{Overview of \codename.}
\label{fig:overviewgenart}
\end{figure}

\kona{This section outlines \codename, \crviv{our} hardware-based design for \codealg. Fig. \ref{fig:overviewgenart} shows \codename's high-level design. It comprises of an individual block for each step of \codealg's online read mapping pipeline: i) \textit{Partitioned Seeding Module} for the Partitioned Seeding step, ii) \textit{Near Memory Seed Locator (NMSL) Module} for the SeedMap Query \crviv{step}, iii) \textit{Paired-Adjacency Filtering Module} for the Paired-Adjacency Filtering step and iv) \textit{Light Alignment Module} for the Light Alignment step.}

\subsection{Partitioned Seeding Module}
This module implements the Partitioned Seeding of the \codealg algorithm (\cref{sec:algo:partitioned_seeding}).
The design objective is to maximize the number of seeds extracted per second. 
Since each hash computation is independent, we compute the hash values of the six seeds of each read pair in parallel, within a single Partitioned Seeding module.
Each hash value calculation is calculated within a single hashing unit that implements \kona{xxHash~\cite{xxhash2024}}. \crviv{The hashing unit implements hashing as a pipelined process to maximize throughput.
This unit is then replicated six times within the Partitioned Seeding module to achieve parallelism, i.e., one module is dedicated for each of the six seeds of the read pair.}

\subsection{Near Memory Seed Locator (NMSL) Module}\label{sec:nmsl}
\label{sec:hw_considerations}

The SeedMap Query stage of \codealg aims to retrieve all the locations of a given seed in the reference genome. 
To optimize the performance of the SeedMap Query stage, we introduce the Near Memory Seed Locator (NMSL), 
i.e., we place the Seed Locator near High Bandwidth Memory (HBM) to benefit from its significant parallelism capabilities~\cite{10.1145/3240302.3240315, singh2021fpga}. Note that NMSL's design is not constrained to HBM. We present a comprehensive discussion of implementation considerations across different memory technologies in \S\ref{sec:mem_tech}.

\begin{figure}[h]
\centering
\includegraphics[width=1.0\columnwidth]{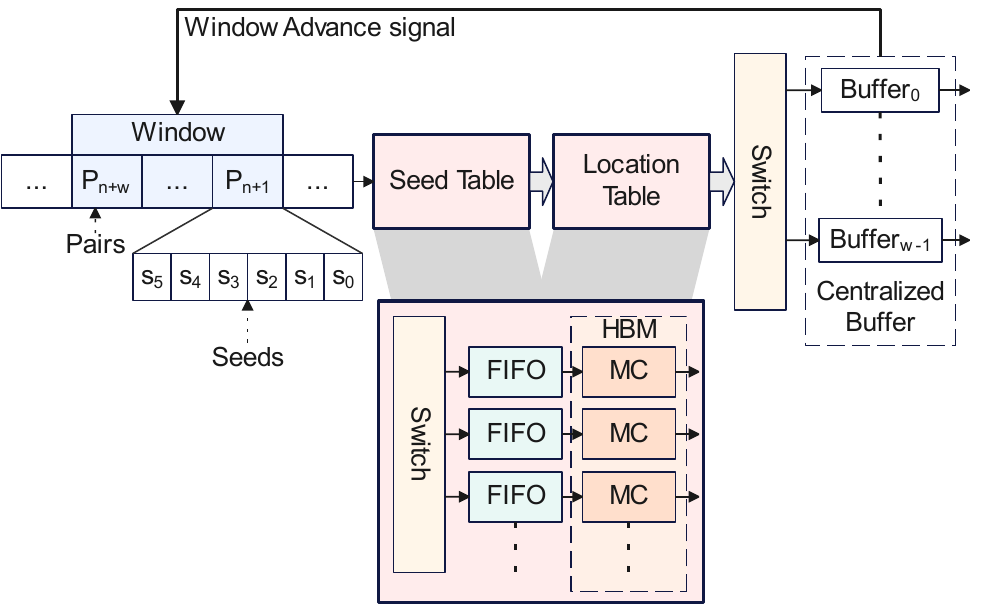}
\caption{Overview of NMSL.}
\label{fig:nearmemory}
\end{figure}

\mypar{Maximizing memory bandwidth} 
To optimize the throughput of NMSL and fully \kona{leverage} HBM parallelism, we adopt a data placement strategy that utilizes all available HBM channels.
To achieve this, we leverage the uniform memory access distribution to the seed and location tables in \codealg. Consequently, we partition these tables into subtables of approximately equal size, ensuring each table fits within the capacity of a single memory channel. These subtables are then allocated across separate memory channels. \CR{Fig.}~\ref{fig:nearmemory} illustrates the partitioning of the tables and the integration of the other components within NMSL. 
Despite the uniform distribution of the seeds and their respective locations across all memory channels, during brief intervals, some channels may become overutilized. To address this imbalance, we introduce FIFOs at the input of each memory channel, ensuring a more balanced distribution of the workload across the channels. This design is generalizable to other memory types, and facilitates simultaneous querying of the seed and location tables while leveraging all channels by using distinct seeds derived from different reads.

\mypar{Handling seed location reordering}
Since we utilize all available memory channels by dispatching multiple reads concurrently, the corresponding responses of six seed locations may arrive out of order. A higher number of channels, such as in HBM, increases the degree of out of order responses. However, the Paired-Adjacency Filtering module (described in \S\ref{sec:paf_hw}) requires all six seed locations in a read-pair to be available simultaneously before its processing can begin. This can result in situations where all allocated Paired-Adjacency Filtering instances are occupied with specific reads and are waiting for their respective straggler seed locations from the HBM. If a response from a new read, which has not yet been assigned a Paired-Adjacency Filtering instance is received, it either has to be discarded or risk causing a deadlock.

\begin{figure*}
\centering
\captionsetup{farskip=2pt,captionskip=-10pt,skip=2pt}
  \includegraphics[width=1.0\textwidth]{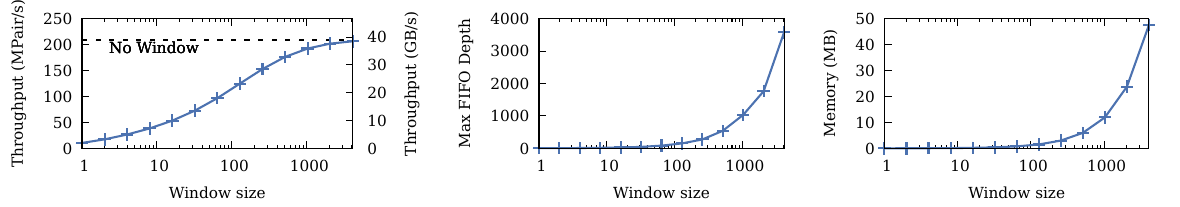} 
\subfloat[\label{fig:window_a}Throughput.]{\hspace{.37\linewidth}}
\subfloat[\label{fig:window_b}Required FIFOs depth.]{\hspace{.35\linewidth}}
\subfloat[\label{fig:window_c}Overall required SRAM.]{\hspace{.3\linewidth}}
\vspace{0.5em}
  \caption{Throughput and hardware costs as functions of the sliding window size.}
  \label{fig:window}
\end{figure*}

To address this issue, we implement a read-pair granularity sliding window mechanism to limit the number of in-flight read-pairs, along with a centralized buffer at the output of the location table, as shown in \CR{Fig.}~\ref{fig:nearmemory}. The sliding window advances by one read-pair after all locations for a given read-pair have been retrieved from the memory and stored in the centralized buffer. Simultaneously, the dispatcher begins forwarding the completely retrieved read-pair locations to the Paired-Adjacency Filtering modules.

The centralized buffer is sized to store the genome locations of read-pairs equal to the window size. This allows NMSL to process multiple read-pairs concurrently while holding the genome locations of the seeds in a read-pair until all six seed locations are received, preventing deadlocks. The buffer is implemented using  FIFOs, each with dual-port SRAM, where each FIFO temporarily stores the genome location of one seed. The total number of FIFOs is the product of the seeds per read-pair (six) and the window size, ensuring all in-flight genome locations fit within the buffer.

In some cases, certain seeds may be associated with an unusually large number of genome locations, requiring excessively large FIFO depths. To address this, \crviii{we leverage an \textit{index filtering threshold},  i.e., we filter out seeds from SeedMap whose number of mapping locations exceed the index filtering threshold. We set the \textit{index filtering threshold} to 500 and thereby also limit the FIFO depth in the centralized buffer to 500. The impact of the \textit{index filtering threshold} on overall accuracy is evaluated in  \S\ref{sec:accuracy}} and \S\ref{sec:indexthr}. Minimap2 employs similar filtering heuristics during the index construction process~\cite{li2018minimap2}.

\mypar{Read-pair sliding window size} The size of the read-pair sliding window influences both the throughput of the design and the hardware costs of the FIFOs and the centralized buffer. We explore this trade-off by simulating NMSL with HBM2 for varying window sizes, using Ramulator 2.0~\cite{luo2023ramulator2}. \CR{Fig.}~\ref{fig:window_a} illustrates the overall throughput of NMSL as a function of the window size, with \CR{Fig.}~\ref{fig:window_b} and \CR{Fig.}~\ref{fig:window_c} depicting the required FIFO depth and the total required SRAM, respectively. The total SRAM requirement is computed as the sum of the SRAM needed for the centralized buffer and the FIFOs.
With a window size of 1024 read-pairs, we achieve 91.8\% of the asymptotic throughput. This window size also constrains the SRAM usage for the FIFO and centralized buffer to an attainable 11.93~MB.

\subsection{Paired-Adjacency Filtering Module}

\label{sec:paf_hw}
The design objective of this module is to maximize the throughput of processed genome locations returned by the \kona{SeedMap Query step}. The Paired-Adjacency Filtering module iterates over the list of genome locations for the first read in the read-pair (read 1) while simultaneously comparing them to the list of locations for the second read (read 2). We implement these lists as hardware FIFOs using dual-port SRAM for both reads. At each cycle, a location from read 1 is compared to a location from read 2. Based on the comparison result, we either read the next location from FIFO1 (for read 1) or from FIFO2 (for read 2) in the subsequent cycle. The two genome locations are output only if the distance between them is smaller than the specified threshold~$\Delta$.

This module's primary logic consists of a comparator used to compare the genome locations. As a result, its performance is mainly constrained by the SRAM used for the FIFOs.

\subsection{Light Alignment Module}
The goal of this module is to efficiently compute the alignment score and identify the positions of edits between two sequences. It implements the algorithm outlined in  \S\ref{sec:light_align}, with two key hardware optimizations designed to accelerate execution: (1) All Hamming masks are computed in a single clock cycle by comparing the read-pairs with shifted copies of the reference, and (2) the start and end segment lengths are computed in parallel for all Hamming masks. Consequently, the number of clock cycles required to compute a single alignment is approximately equal to the length of the sequences being compared.

\section{Evaluation Methodology}
\label{sec:methodology}

\crviii{We evaluate \codename in terms of  performance, area, and accuracy. We first present the performance, area and power analysis of individual steps and demonstrate our methodology to create a balanced, high-throughput design. We then proceed to compare \codename with software and hardware-based state-of-the-art read mapping workflows.}
\crviii{We compare the throughput per unit area (Mbp/s/mm\textsuperscript{2}) and throughput per unit power (Mbp/s/W) of \codename{} against the configurations presented in the beginning of \S\ref{sec:methodology}. Throughput is measured in Mbp/s to ensure a fair comparison with GenCache~\cite{nag2019gencache} where throughput was evaluated with single-end reads of size 100\,bp and expressed in KReads/s. For our evaluation, we convert this throughput into Mbp/s.}

\noindent\textbf{Evaluated Systems}. We perform an end-to-end evaluation of read mapping across software baselines, hardware-accelerated baselines, and our proposed \kona{systems}.\\
\kona{\textbf{{\small Software Baseline.}}}
(i) Minimap2 (MM2)~\cite{li2018minimap2}: A state-of-the-art software read mapper executed on a general-purpose CPU. \\
\textbf{{\small Hardware-accelerated baselines.}} 
(ii) BWA-MEM-GPU~\cite{10.1145/3577193.3593703}: An end-to-end GPU implementation of \crviii{BWA-MEM~\cite{10.1145/3577193.3593703}}. For this baseline, we use the results reported for an NVIDIA A100 GPU.
(iii) GenDP~\cite{gu2023gendp}: A state-of-the-art hardware framework for accelerating dynamic programming, used for running Minimap2. In our evaluation, we assume an ASIC implementation of GenDP that accelerates Minimap2's chaining and alignment stages. 
(iv) GenCache~\cite{nag2019gencache}: A hardware-accelerated read mapper optimized for short single-end reads. Area, power, and throughput are taken from the values reported in ~\cite{nag2019gencache}.\\
\textbf{{\small Our proposed \kona{systems}.}} (v) \kona{\codealg + MM2}: \crviii{Our software pipeline, i.e., \codealg,} used alongside Minimap2 on general-purpose CPUs to improve performance. Reads not mapped by \codealg's Paired-Adjacency Filtering or Light Alignment stages are handled by Minimap2. (vi) \kona{\codename + GenDP}: Our ASIC accelerator of \codealg is integrated with GenDP, where GenDP serves as the fallback for reads that \codename cannot map (see \CR{Fig.}~\ref{fig:fallback}).

\mypar{\crviii{CPU, GPU and HBM configurations}}
\crviv{For the evaluation of software-based and GPU-based configurations,} 
we utilize the hardware specified in \CR{Table}~\ref{table:baseline}.
CPU power consumption is measured using Intel RAPL~\cite{guide2011intel}, and 
GPU power is measured using Nvidia's nvprof profiler~\cite{nvprof}.
HBM timing and power consumption are modeled using \crviii{Ramulator2~\cite{luo2023ramulator2,ramulatorsource2,ramulatorsource,kim2016ramulator}} and  DRAMsim3~\cite{8999595}, respectively. Power consumption is estimated by feeding Ramulator2-generated traces into DRAMsim3. Both simulators are configured for \crviii{HBM2e~\cite{hbm2e}}, with each stack built from eight 8\,Gb dies (8\,GB per stack). We use four stacks for a total of 32\,GB. Each stack exposes eight 128-bit channels (1\,GB per channel; 1024-bit aggregate per stack). The HBM runs at 1\,GHz in DDR mode, i.e., 2\,GB/s per data pin.

\begin{table}[h]
\caption{\crviii{CPU and GPU} \crviv{configurations}.}
\vspace{-0.5em}
\scriptsize
\begin{minipage}[t]{.55\linewidth}
\vspace{0pt}
\begin{center}
\begin{tabular}{@{}rl@{}}
\toprule
\multicolumn{2}{c}{\textbf{Intel Xeon Gold 6238T}} \\
\midrule
Cores & 22 @ 1.9\,GHz\\
Cache & L1 D\&I 22$\times$32\,KB, 
22$\times$32\,KB\\
&L2 22$\times$1\,MB, Shared L3 30.25\,MB\\
Memory & 96\,GB DDR4
6 Ch. 2933\,MT/s  \\
Die Area & 300\,mm\textsuperscript{2} \\
 \bottomrule 
\end{tabular}
 \end{center}
  \end{minipage}%
    \begin{minipage}[t]{.45\linewidth}
    \vspace{0pt}
    \begin{center}
    \begin{tabular}{@{}rl@{}}
 \toprule
\multicolumn{2}{c}{\textbf{NVIDIA  Quadro GV100}}\\
\midrule
Cores & 5120 @ 1.6\,GHz \\
Cache & L2 6\,MB \\
Memory & 32\,GB HBM2 \\
Die Area & 815\,mm\textsuperscript{2} \\
\bottomrule
\end{tabular}
 \end{center}
 \end{minipage} 

\label{table:baseline}
\end{table}

\mypar{Evaluation of hardware cost}
Partitioned Seeding, NMSL, Paired-Adjacency Filtering, and Light Alignment are implemented in SystemVerilog\crviii{~\cite{SystemVerilogIEEE}}. We synthesize and place-and-route these blocks in a commercial 28\,nm CMOS technology\crviii{~\cite{iwai1999cmos}}. All components operate at 2.0\,GHz. SRAM area and power are modeled with CACTI 7.0~\cite{cacti, 10.1145/3085572} using a 22\,nm process configuration and a 2.0\,GHz timing target. \crviii{For fair comparison with GenDP~\cite{gu2023gendp}, \crviv{which} reports values \crviii{at} 7\,nm, the area and power of the CPU, GPU, GenCache~\cite{nag2019gencache}, and \codename components are also scaled to a 7\,nm process~\cite{STILLMAKER201774}}.

\mypar{Datasets}
We use the complete human reference genome GRCh38~\cite{giab_grch38} to perform all  experiments. We build SeedMap with the human reference genome and apply a filter threshold of 500 for the maximum number of locations, as described in \S\ref{sec:nmsl}.
For input read datasets, we use real human paired-end read sets from Ashkenazi Son HG002 (NA24385) of size 150\,bp provided by the \emph{Genome in a Bottle} Project~\cite{zook2016extensive, giab_hg002} in all the experiments. We use a 1\,M pairs subset to evaluate the throughput and latency of all \codename{} components (\S\ref{sec:sizing-balancing}), and 100\,M pairs for the end-to-end throughput evaluation (\S\ref{sec:e_2_e}). For long reads, we use a PacBio HiFi real human dataset from \crviii{~\cite{giab_hg002}}, containing 222 million reads with an average read length of 9,569 bp.

\mypar{Accuracy evaluations}
We map a read set with a coverage depth of 100$\times$---on average, each base of the genome is covered by 100 reads---and produce BAM alignment files~\footnote{BAM\crviii{~\cite{Li2009SAM}} is the compressed binary format storing read-to-reference alignments and related metadata}. We then use the \emph{freebayes}~\cite{garrison2012haplotypebasedvariantdetectionshortread} tool to call single-nucleotide polymorphisms (SNP) and insertions/deletions (INDEL) from the BAM files for both our method and all baselines. After variant calling, we use \emph{vcfdist}\crviv{\cite{dunn2023vcfdist}} to compare the resulting SNPs and INDELs against the values in the Genome in a Bottle variant set~\cite{zook2016extensive, giab_hg002,giab_benchmark_hg002}, and then report standard accuracy metrics. Variants absent from the Genome in a Bottle benchmark set are counted as false positives, and benchmark variants not recovered are false negatives.

\section{Evaluation}
\subsection{\crviii{Near Memory Seed Locator Module Performance Evaluation}} \label{sec:nmsl_eval}

\begin{figure}[h]
    \centering
     \includegraphics[width=\linewidth]{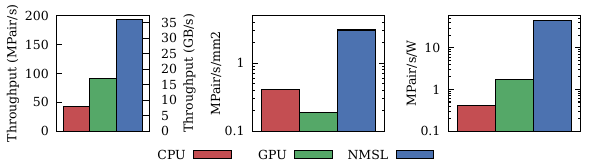}
    \caption{ \crviii{Throughput of the SeedMap Query stage implemented by NMSL Module (higher is better)}.}
    \label{fig:HBM_throughput}
\end{figure}

\crviv{The throughput of the Near Memory Seed Locator Module (\S\ref{sec:nmsl}) implementing the SeedMap Query step (\S\ref{sec:algo:seedmapquery})} determines the throughput of the entire GenPairX design. While the throughput of the compute units can be \crviii{relatively} easily scaled by replicating the hardware, the maximum throughput of the NMSL unit is fundamentally limited by the bandwidth available from the HBM and the number of accessible HBM channels. We evaluate the performance of NMSL and 
compare it with two alternative configurations: (i) \textbf{GPU-based NMSL implementation}: we develop an optimized CUDA kernel to perform SeedMap query lookups. \crviii{We \crviv{use} the same optimized data structure for SeedMap presented in \S\ref{sec:seedmap} to ensure parallelism}, and (ii) \textbf{CPU-based NMSL implementation}: we use a multi-threaded implementation, with each thread repeatedly executing the SeedMap lookup logic. We apply the same approach as outlined in \S\ref{sec:nmsl} to achieve the maximum bandwidth for both DDR and HBM in CPU- and GPU-based implementations, respectively.
We observe the following key performance characteristics based on the results \crviii{reported in} \CR{Fig.}~\ref{fig:HBM_throughput}: (1) NMSL achieves a 2.12$\times$ throughput improvement over the GPU, despite both systems utilizing HBM2 with 32 memory channels. This performance gap arises from NMSL's more efficient and specialized architecture, which better utilizes \crviii{memory} bandwidth compared to the GPU's traditional synchronization and caching mechanisms. The GPU also suffers from warp divergence due to the variability of the workload, as observed in prior studies~\cite{gu2023gendp}. (2) NMSL achieves 16.1$\times$ higher throughput per unit area and 26.8$\times$ higher throughput per unit power than GPU. This improved efficiency is primarily due to the GPU’s \crviii{many} compute elements, which are not required for our table lookup mechanism, and data movement across the GPU memory hierarchy. (3) Compared to the CPU-based implementation with DDR5 memory channels, NMSL \crviii{provides} a 4.58$\times$ speedup, due to custom hardware that effectively utilizes memory-level parallelism provided by HBM. 

\subsection{GenPairX  Design: Sizing and Balancing} \label{sec:sizing-balancing}

The maximum sustained throughput of NMSL determines the overall throughput of \codename{}. This throughput value is used to determine the number of hardware instances required for each of \codename{}'s hardware modules, as detailed in \CR{Table}~\ref{tab:breakdown}. This section calculates the individual execution time of each module and leverages the results to design and size a balanced pipeline.
Note that the hardware NMSL module achieves a sustained throughput of 192.7\,MPair/s, as seen in \crviii{\S\ref{sec:nmsl_eval}, Fig.\ref{fig:HBM_throughput} (left)}. 

\begin{table}[h]
\caption{\codename{} modules.}
\vspace{-1.5em}
\begin{center}
\footnotesize
\begin{tabular}{@{}rrrr@{}}
\toprule
\textbf{Modules} & \textbf{Throughput} &\textbf{Latency} & \textbf{\# Instances}\\ 
&\textbf{[MPair/s]}&\textbf{\# Cycles} \\
\midrule
Partitioned Seeding & 333\hspace{1.88mm} & 10\hspace{1.88mm} & 1 \\
Paired-Adjacency Filtering & 83.0\hspace{1.88mm} & 24.1\hspace{1.88mm}& 3 \\
Light Alignment & 1.1\hspace{1.88mm}\hspace{1.88mm}&  156 \hspace{1.88mm}&  174 \\
\bottomrule
\end{tabular}
\label{tab:breakdown}
\end{center}
\end{table}

\noindent\textbf{Partitioned Seeding module.} The number of cycles required per input are independent of the input data. A single instance of the Partitioned Seeding module processes 333\,MPair/s, which exceeds the throughput of NMSL. As a result, only one instance of the Partitioned Seeding module is necessary.

\mypar{Paired-Adjacency Filtering module} \crviii{The number of cycles required to process a single read-pair in the Paired-Adjacency Filtering module varies based on the number of reference locations returned by NMSL.
We leverage profiling results of our CPU-based \codealg{} implementation to acquire 
the average number of filtering iterations needed. \crviv{Based on this, we calculate a throughput requirement of 83\,MPair/s for the Paired-Adjacency Filtering module and therefore we leverage 3 instances of the Paired-Adjacency Filtering module to match the throughput of the pipeline.}}

\noindent\textbf{Light Alignment module.} The number of cycles required per input is independent of the input data. 
We calculate the throughput based on the execution cycles required by a single alignment and the average number of alignments required by a read-pair. 
The module processes one alignment in 156 clock cycles. Specifically, it computes eight consecutive Hamming masks in a single clock cycle, traverses the masks from both ends over the $read\_length$ cycles to identify the longest continuous segment of `1's, and compares segment lengths to determine the longest one. Based on our profiling \CR{in \S\ref{sec:analysis}}, 11.6 alignments are needed for a single read-pair on average. Consequently, a single Light Alignment instance can process 1.1\,MPair/s, and 174 instances are required to support NMSL's throughput. 

\noindent\textbf{Optimization for Balancing.} 
\crviii{Potential congestion issues may arise for reads with seed mapping locations that exceed the average number \crviv{seen during} profiling in \S\ref{sec:obs2}.}
To address this challenge and prevent the stalling of the entire pipeline, we introduce a circular buffer implemented in SRAM, positioned immediately before the \crviii{Light Alignment} modules as well as between the NMSL and the Paired-Adjacency Filtering modules. This buffer helps \crviii{to balance} the workload and maintains consistent performance.

\subsection{Area and Power Analysis} \label{sec:area_power}
Based on \CR{Table}~\ref{tab:breakdown}, we measure the area and power consumption of the total required instances per hardware module. The FIFOs before the seed and location tables and the centralized buffer sizes correspond to the requirements detailed in \CR{\S\ref{sec:hw_considerations}}. We compute the area and power of the HBM PHY based on existing chips~\cite{hwang2020hotchips, locuza_nvidia_ada}, shown in \CR{Table}~\ref{tab:area_power}.

\begin{table}[h]
\begin{threeparttable}
\caption{Area and power breakdown of \codename+GenDP.}
\footnotesize
\begin{tabularx}{\columnwidth}{@{}rrr@{}}
\toprule
\textbf{Components} & \textbf{Area [mm\textsuperscript{2}]} & \textbf{Power [mW]} \\
\midrule
Partitioned Seeding\tnote{a}  & 0.016 & 82.4 \\ 
Paired-Adjacency Filtering\tnote{a} & 0.027 & 15.6\\ 
Light Alignment\tnote{a} & 0.53 & 453.6 \\ 
HBM PHY & 60.0 & 320.0\\
Centralized Buffer (11.74\,MB)\tnote{b} & 6.13 & 6.09 \\
FIFOs (190\,KB)\tnote{b}  & 0.091 & 3.36\\
\midrule
\textbf{\codename{}} & \textbf{66.80} & \textbf{881.05}\\
\midrule
GenDP Chain \cite{gu2023gendp}  & 174.9& 115.8$\times 10^{3}$\\
GenDP Align \cite{gu2023gendp}  & 139.4& 92.3$\times 10^{3}$ \\
\midrule
\textbf{\codename{} + GenDP} & 381.1 & 209.0$\times 10^{3}$\\ 
\bottomrule
\end{tabularx}
\label{tab:area_power}
\begin{tablenotes}
\footnotesize
\item[a]Synthesis/Place-and-Route and Power Sim in 28nm, scaled with power and area scaling factor 3.5 and 1.91 (20$\rightarrow$7) from Stiller et al.\cite{STILLMAKER201774}
\item[b] Numbers generated with CACTI 7.0 for 22nm \cite{cacti}, and scaled with power and area scaling factor 3.5 and 1.91 (20$\rightarrow$7) from Stiller et al.\cite{STILLMAKER201774}
\end{tablenotes}
\end{threeparttable}
\end{table}

\subsection{GenDP Integration \& End-to-End Analysis}
\label{sec:e_2_e}

\begin{figure}[h]
\centering
\includegraphics[width=1.0\columnwidth]{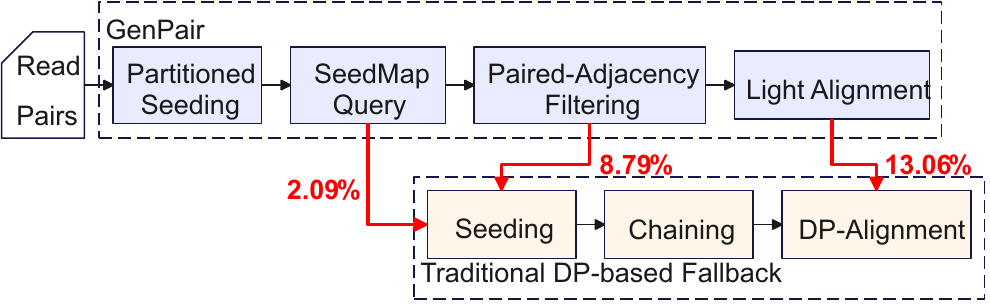}
\caption{\kona{Residual read pairs that cannot be mapped or aligned by \codealg.}}
\label{fig:fallback}
\end{figure}

\crviv{\mypar{Residual read pairs}
\codealg's pipeline results \crviii{in} a small percentage of residual read pairs that cannot be directly mapped.
Fig.~\ref{fig:fallback} illustrates the stages of \codealg and the percentage of read pairs requiring fallback to \crviii{traditional DP-based alignment methods}. In the SeedMap Query stage, 2.09\% of reads fail to match any hash table entries and revert to traditional DP-pipeline. During Paired-Adjacency Filtering, 8.79\% of read pairs exceed the positional distance threshold $\Delta$ and fall back to traditional DP-pipeline. In the Light Alignment stage, 13.06\% of read pairs are not aligned but instead undergo direct DP alignment. These bypass seeding and chaining, as their potential mapping locations are already identified.}

We evaluate the impact of \codename{} on an end-to-end
read alignment use case, which accounts for the residual read-pairs that cannot be directly mapped by \codename{}, but need traditional DP-alignment.
For this purpose, we consider the \emph{\codename{} + GenDP} system, where GenDP~\cite{gu2023gendp} is a hardware framework designed to accelerate chaining and DP-Alignment using the \crviii{Banded Smith-Waterman algorithm\cite{chao1992aligning,harris2007banded}}.

\mypar{GenDP integration}
We evaluate the bandwidth requirements for residual chaining and alignment tasks by quantifying the throughput in  ``Million Cell Updates Per Second'' (MCUPS), which represents the number of values computed per second in GenDP’s DP-Matrix. Analyzing a dataset of 1 million reads, we find that 331,772 MCUPS are needed for residual chaining, and 3,469,180 MCUPS for alignment. Using GenDP’s area cost (MCUPS/mm\textsuperscript{2}) and power consumption (MCUPS/mW), we calculate the area and power requirements. The estimated area and power values for GenDP to match our design's residual  throughput  are presented in \CR{Table}~\ref{tab:area_power}.

\codename{} and GenDP are interconnected through an on-chip \crviii{AXI-Stream\cite{ARM_AXI4Stream}} communication bus. For a 7\,nm CMOS technology node, the area overhead of the communication bus is 1\,mm\textsuperscript{2}, with power consumption of 50 mW.
To manage potential congestion during burst traffic periods, additional FIFOs are introduced. These SRAM FIFOs use 2-bit encoding, allowing a read-pair to be represented with 75 bytes.
\crviv{For a batch of 10K reads, the area overhead of the FIFO is 1.3\,mm\textsuperscript{2}, with a power consumption of 500\,mW.}
%
\crviii{Thus,} the area and power consumption for the interconnect and the FIFOs are negligible in the context of the overall design.

\mypar{Host integration} 
\crviii{As explained in \S\ref{sec:nmsl_eval}, \codename{} saturates when the input throughput is 192.7\,MPair/s. Given a 2-bit encoding and a read length of 150\,bp, the host must provide 14.5\,GB/s \crviv{memory} bandwidth \crviv{for input data transfer}. The output includes the locations (8 bytes) and CIGAR strings ($\sim$20 bytes) for each pair. Therefore 5.4\,GB/s bandwidth is required. Transfer time is not a limiting factor, as bandwidth is the primary concern, and there are no data dependencies between read and write operations. Both 16-lane \crviii{PCIe Gen 3\cite{PCIeGen3} and 4\cite{PCIeGen4}} interfaces support these bandwidth requirements.}

\mypar{End-to-end performance evaluation}
\crviii{We compare \codename{}+GenDP against the evaluated systems MM2, BWA-MEM-GPU, GenCache, GenDP and GenPair+MM2 presented in the beginning of \S\ref{sec:methodology} in terms of \emph{throughput per unit area} (Mbp/s/mm\textsuperscript{2}) and \emph{throughput per unit power} (Mbp/s/W).
\crviv{We evaluate GenCache with single-end reads of size 100\,bp. We measure throughput in KReads/s}.}
\crviii{For a fair comparison of \codename{}+GenDP with GenCache, we also convert GenCache's KReads/s to Mbp/s.}
\crviii{We obtained the results of BWA-MEM GPU as reported in prior works, specifically executed on the A100 GPU~\cite{10.1145/3577193.3593703}.}

\begin{figure}[h]
    \centering
     \includegraphics[width=\linewidth]{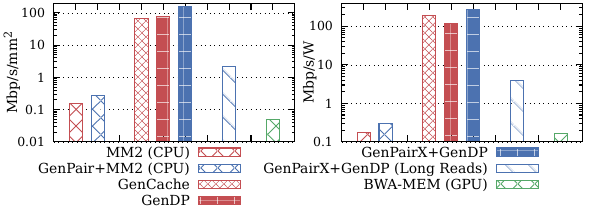}
    \caption{\crviii{Performance per unit area (left) and performance per power (right). Higher is better.}}
    \label{fig:throughput_e2e}
\end{figure}

\crviv{Based on Fig.~\ref{fig:throughput_e2e}, we make six observations. First, \codename{} + GenDP has 1.97$\times$ and 2.38$\times$ higher performance per unit area and performance per unit power respectively compared to GenDP. This improvement is due to \codename{}'s ability to alleviate the majority of compute-intensive alignment and chaining operations, which significantly reduces power consumption and area.}
\crviv{Second, compared to the single-end read mapper GenCache, \codename{} + GenDP provides 2.35$\times$ and 1.43$\times$ improvements in area and power, respectively. This demonstrates \codename{}’s ability to process more complex reads, which carry richer sequence information, at higher throughput than state-of-the-art single-end hardware read mapper.
Third, \codename{} + GenDP outperforms BWA-MEM GPU by a factor of 3053$\times$ in performance per unit area and 1685$\times$ in  performance per unit power, due to the low SIMD utilization observed during the seeding stage (49\%).}

\crviv{Fourth, \codename{} + GenDP achieves better performance per unit area and power compared to the software baselines. Specifically, compared to the CPU-based \kona{GenPair} + MM2 implementation, \codename{} + GenDP provides 557$\times$ better performance per unit area and 911$\times$ better performance per unit power. \codename{} + GenDP provides 958$\times$ better performance per unit area and 1575$\times$ better performance per unit power compared to MM2.}

\crviv{Fifth, Fig.~\ref{fig:throughput_e2e} highlights the algorithmic efficiency of \codealg, as \codealg + MM2 provides 1.72$\times$ higher performance than MM2.} 
\crviii{\crviv{Sixth}, when applied to long reads using the methods described in \S~\ref{sec:longreads}, \codename{} achieves a throughput roughly one order of magnitude lower than for short reads. We observe the same trend for the software read mapper Minimap2. This is because long reads, due to their higher noise and complexity, require more DP fallback. In our simulator, we find that the DP-alignment step constitutes the main bottleneck.}

\crviii{For a straightforward comparison, \CR{Table}~\ref{tab:absolute} shows the absolute values for the performance of GenDP, GenCache and \codename{} + GenDP. \codename{} + GenDP achieves 26.6$\times$ and 2.4$\times$ higher throughput than GenCache and GenDP, respectively.}
\begin{table}[h]
\caption{Absolute performance of hardware accelerators.}
\vspace{-1.5em}
\begin{center}
\footnotesize
\begin{tabular}{rrrr}
\toprule
\textbf{HW accelerator} & \textbf{Area [mm\textsuperscript{2}]} & \textbf{Power [W]} & \textbf{Tput [Mbp/s]} \\
\midrule
GenCache & 33.7 & 11.2 & 2,172 \\
GenDP & 315.8 & 209.1 & 24,300\\
\codename{} + GenDP & 381.1 & 209.0 & 57,810 \\
\bottomrule
\end{tabular}
\label{tab:absolute}
\end{center}
\end{table}

\subsection{\codename{} \crviii{Scalability Across Memory Technologies}} \label{sec:mem_tech}
To evaluate \codename{}'s scalability across different memory technologies, we analyze SeedMap \kona{Query} throughput, required PEs, and power consumption for three memory configurations. \CR{Table}~\ref{tab:memtype} presents the absolute throughput and throughput per unit power, highlighting two key findings: (1) \codename{}'s throughput scales with the available \crviii{memory} bandwidth, which depends on the number of memory channels and the technology. \crviii{For example, when using HBM2, throughput increases by 11.4$\times$ and 9.8$\times$, respectively, in comparison to using DDR5 and GDDR6. (2) Throughput per watt demonstrates less variance (0.04\%-0.16\%) in comparison to the throughput fluctuation \crviv{across} different memory technologies.}
This is due to GenDP being the dominant power consumer in \codename{}.  

\begin{table}[h]
\caption{Memory technology comparison.}
\vspace{-1.5em}
\begin{center}
\footnotesize
\begin{tabular}{@{}rcc@{}}
\toprule
\textbf{Memory Type} & \textbf{Throughput} &\textbf{Throuput per unit power}\\ 
&\textbf{[MPair/s]}&\textbf{[MPair/s/W]} \\
\midrule
DDR5 (4 channels) &
16.91\hspace{1.88mm}\hspace{1.88mm}&0.75\hspace{1.88mm}\\
GDDR6 (8 Channels) & 19.80\hspace{1.88mm}\hspace{1.88mm}&  0.79\hspace{1.88mm}\\
HBM2 (32 Channels) & 192.7\hspace{1.88mm}\hspace{1.88mm}&  0.91\hspace{1.88mm}\\
\bottomrule
\end{tabular}

\label{tab:memtype}
\end{center}
\end{table}

\subsection{Accuracy Analysis}
\label{sec:accuracy}
The heuristics employed in the \codealg algorithm may, in some edge cases, produce results that differ from those of MM2. 
As mentioned in \S\ref{sec:nmsl}, \codealg \crviv{ filters out seeds with more than 500 mapping locations from SeedMap}. To quantify the impact of the filter on the mapping accuracy, we assess \codealg both with and without the filter. We perform variant calling and compare the results \crviii{using} a recent genomic variation benchmark~\cite{Wagner2022}. We repeat the same experiment with MM2 for comparison.

The variant calling results for MM2 and GenPair + MM2 (with and without the filter) are in \CR{Table}~\ref{tab:vc}. This tables shows the number of true positives (TP), false positives (FP), precision, and $F_1$ score for three mappers. We make the following observations: (1) \crviii{Although \codealg + MM2 leverages lightweight heuristics compared to MM2, the $F_1$ score reduces only by 0.0026 for SNPs and INDELs}. This difference is negligible based on prior benchmarking studies of variant calling pipelines~\cite{Barbitoff2022}.
(2) \kona{GenPair} + MM2 (both with and without the filter) demonstrates better overall precision than MM2. This improvement is due to a lower number of false-positive variants detected by \kona{GenPair} + MM2.
(3) The filter has negligible impact on accuracy. \crviii{The use of the filter in \kona{GenPair} + MM2 results in an $F_1$ score reduction of 0.0001 for INDELs and no reduction for SNPs}.  This minimal difference likely arises from cases where the optimal mapping location is filtered out, but \codealg still maps the read to a suboptimal location. 

\begin{table}[h]
\caption{Variant calling benchmark results.}
\vspace{-1.5em}
\begin{center}
\footnotesize
\begin{tabular}{@{}ll@{} 
 r@{ }r@{\  }r@{\  }r@{\  }r@{}}
\toprule
&  & \textbf{TP (\#)} & \textbf{FP (\#)}  & \textbf{Prec.} & \textbf{Rec.} & $\mathbf{F_1}$ \\
\midrule
\multirow {3}{*}{SNP} & MM2 & \textbf{2390698} & 2440 & 0.9989 & \textbf{0.9954} & \textbf{0.9913} \\
&  \kona{GenPair}+MM2 no filter\ \ \  & 2378846 & \textbf{994} & \textbf{0.9996} & 0.9939 &0.9887 \\
& \kona{GenPair}+MM2 & 2381480 & 1040 & 0.9996 & 0.9938 & 0.9887 \\
\midrule
\multirow {3}{*}{INDEL} & MM2 & \textbf{596036} & 7334 & 0.9878 & \textbf{0.9597} &\textbf{0.9326} \\
&  \kona{GenPair}+MM2 no filter & 589962 & \textbf{7088} & \textbf{0.9881} & 0.9583 & 0.9300 \\
\textit{}& \kona{GenPair}+MM2 & 584616 & 7102 & 0.9880 & 0.9582 &  0.9299\\
\bottomrule
\end{tabular}
\label{tab:vc}
\end{center}
\end{table}

\subsection{Sensitivity Analysis to Read Error Rate}

To evaluate the applicability of our approach across different applications that can use datasets with varying error rates, we use Mason~\cite{Holtgrewe2010Mason} to generate synthetic read pairs from the full human reference genome with different per-base error rates. We use Mason’s default error profile that uniformly distributes errors \crviii{across} substitutions, insertions, and deletions.

\begin{figure}[h]
\captionsetup{farskip=2pt,captionskip=-5pt,skip=5pt}
\centering
\includegraphics[width=1.0\columnwidth]
{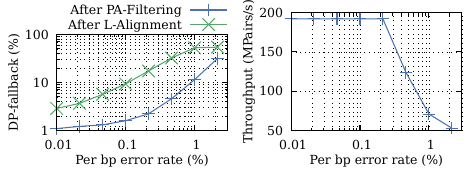}
\subfloat[\label{fig:synthetic_a}DP fallback ratio]{\hspace{.5\columnwidth}}
\subfloat[\label{fig:synthetic_b}\codename{} + GenDP throughput.]{\hspace{.5\columnwidth}}
\vspace{-0.5em}
\caption{Impact of error rate on DP fallback and throughput.}
\label{fig:synthtetic}
\end{figure}

\CR{Fig.}~\ref{fig:synthetic_a} shows the percentage of reads that fall back to DP (DP fallback) after Paired-Adjacency Filtering and after Light Alignment. Reads that fall back after Paired-Adjacency Filtering proceed to DP chaining and DP alignment, whereas reads that fall back after Light Alignment require only DP alignment. We make two key observations. First, as the sequencing error rate increases beyond 0.1–0.2\%, a larger fraction of read pairs require DP fallback, \crviii{due to an increase in potential mappings and edit variance}.
Second, under Mason’s default error profile, a greater proportion of reads require DP fallback after Light Alignment than after Paired-Adjacency Filtering.

In \CR{Fig.}~\ref{fig:synthetic_b}, we present the throughput of \codename combined with GenDP. We make two key observations. First, for error rates below 0.2\% per base pair, the throughput remains stable. Our detailed analysis of our simulation results indicates that, at low error rates, the SeedMap Query stage is the bottleneck in the pipeline. Since this stage maintains high efficiency at low error rates, it effectively caps the overall throughput at approximately 192 million pairs per second (MP/s), explaining the observed stability. Second, for error rates above 0.2\% per base pair, the throughput decreases due to the increasing number of read pairs that require DP alignment to be mapped to the reference genome. Specifically, for the synthetic dataset, our simulations reveal that the DP alignment stage becomes the primary bottleneck in the pipeline.

Regardless of whether the input consists of long or short reads, sequencing technologies are becoming increasingly accurate, reaching error rates as low as 0.0005\%~\cite{Cheng2024, Sun2024}. \codename follows this general trend, as its performance remains stable for per-base error rates below 0.2\%. For future sequencing technologies, it may be advantageous to reduce the number of costly DP PEs, since higher read accuracy decreases the need for DP fallback. Such a design choice would significantly reduce both power consumption and area.

\subsection{Sensitivity Analysis of the Index Filtering Threshold}
\label{sec:indexthr}

To evaluate the impact of the index filter threshold \crviii{(see \S\ref{sec:hw_considerations})}, we generate synthetic reads based on the full human reference genome. We use Mason~\cite{Holtgrewe2010Mason} to introduce sequence variations (SNP rate of $10^{-3}$ and INDEL rate of $2 \times 10^{-4}$) and to simulate reads using Mason's default sequencing error profile. We then construct multiple indexes with different filter thresholds and run \codealg (without DP fallback) using the synthetic reads as input and the corresponding indexes. To assess mapping accuracy, we use \textit{paftools}~\cite{li2018minimap2} to compare our mapping results against the baseline, verifying only the correctness of the mapping location rather than the full alignment.

\begin{figure}[h]
\captionsetup{farskip=2pt,captionskip=-5pt,skip=5pt}
\centering
\includegraphics[width=1.0\columnwidth]
{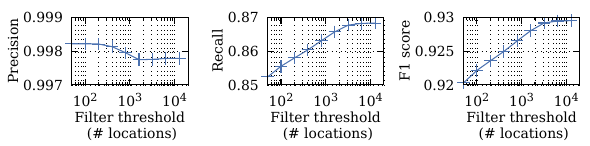}
\caption{Impact of the index filter threshold on the number of mapped pairs and incorrect mappings of \codealg{}}
\label{fig:filter}
\end{figure}

In \CR{Fig.}~\ref{fig:filter}, we show the \emph{precision}, \emph{recall}, and \emph{F1} score obtained from our experiments. We make two key observations. First, as the filter threshold increases, precision decreases while recall increases. This behavior indicates that both the number of mapped pairs and the number of incorrect mappings grow with a higher filter threshold. In other words, a less restrictive filter allows more seeds to pass, resulting in a higher number of mapped read pairs but also an increased number of incorrectly mapped reads. Second, beyond a filtering threshold of 4000, precision, recall, and, consequently the F1 score stabilize.

Based on our observations above, we set the index filtering threshold to 500. This value represents a good trade-off between precision and recall and is also the default configuration in the widely used Minimap2~\cite{10396343}. As shown in \CR{Fig.}~\ref{fig:filter}, there is no single threshold that simultaneously maximizes both precision and recall. A higher threshold increases recall but leads to a notable drop in precision, which can be unacceptable for accuracy-critical applications such as variant detection\crviii{\cite{koboldt2020best,meynert2014variant}}. In such applications, missing even a single DNA variant may have a significant impact~\cite{dunn2023vcfdist}. Conversely, a lower threshold improves precision but reduces recall, meaning that some reads fail to be mapped and the pipeline must rely more heavily on the DP state, negatively affecting performance. Therefore, choosing a threshold of 500 \crviii{strikes a balance between mapping accuracy and performance}.

\section{Related Work}
\crviii{To our knowledge, this is the first work to 1) introduce a read mapping algorithm for \crviv{\emph{paired-end genomic}} reads and 2) present the first hardware-algorithm co-designed accelerator for \crviv{\emph{paired-end}} read mapping. In this section, we briefly review prior work on acceleration of the computationally intensive tasks of read mapping, such as string matching, indexing, and dynamic programming (DP).}

\mypar{FPGA-based paired-end read accelerators} \crviii{Many works design FPGA-based accelerators~\cite{Liyanage2023,6239809,Samarasinghe2021Energy,alser2019shouji,alser2020sneakysnake,alser2017gatekeeper,koliogeorgi2022gandafl,guo2019hardware,singh2021fpga,chen2021high,fei2018fpgasw,haghi2021fpga,rucci2018swifold,li2021pipebsw,liao2018adaptively} to directly accelerate the performance bottlenecks caused by DP operations. Yang et al.~\cite{10177185} propose an FPGA accelerator for BWT-based seed-and-extend mapping, which involves costly index searches. In contrast, \codename{} adopts the seed-chain paradigm and \crviv{leverages specific features of paired-end reads to reduce and accelerate computation}.}

\mypar{End-to-end hardware accelerators for read mapping} Numerous studies have introduced end-to-end hardware accelerators for read mapping\crviii{~\cite{dunn2021squigglefilter,fujiki2018genax, nag2019gencache, turakhia2018darwin,Lindegger2023Scrooge,cali2020genasm,turakhia2018darwin,banerjee2018asap,fujiki2020seedex,cali2022segram,HOUTGAST201854, 6972452}.} For instance, GenAx~\cite{nag2019gencache} and GenCache~\cite{nag2019gencache} accelerate short single-end read processing, while Darwin~\cite{turakhia2018darwin} accelerates long-read sequencing. 
\codename{} is the first hardware accelerator specifically designed for \crviii{\emph{paired-end}} read mapping, leveraging the additional performance opportunities presented with paired-end reads.

\mypar{Pre-alignment filters} 
\crviv{Various} works propose filters~\cite{10.1145/3503222.3507702, alser2020sneakysnake, alser2017gatekeeper,xin2013accelerating,xin2015shifted,kim2018grim,alser2019shouji} for \crviv{\emph{single-end}} reads (i.e., non-paired-end reads) to quickly filter out dissimilar sequences. These filters effectively reduce the workload of fine-grained compute-intensive operations (i.e., DP), thereby increasing the performance of read mapping.
Some of them introduce hardware accelerators and approximate matching methods to reduce the computational cost of pre-alignment filtering algorithms, e.g., GateKeeper~\cite{alser2017gatekeeper}, SneakySnake~\cite{alser2020sneakysnake}, 
\crviv{\emph{Shifted Hamming Distance (SHD)}}~\cite{xin2015shifted} and  GRIM-Filter~\cite{Kim2018}.
\codename{}'s Light Alignment method and SneakySnake both use the SHD approach.
\codename{} always returns the optimal alignment \crviii{given an upper limit for the number of edits, while SneakySnake\cite{alser2020sneakysnake} can handle more edits.}
A combination of the two methods is a promising future work.
GRIM-Filter\cite{kim2018grim} is a Processing-In-Memory(PIM)-based filtering mechanism designed for short (single-end) read mapping, relying on a seeding approach that determines the presence or absence of seeds within predefined bins. \crviii{The filtering mechanism uses a fixed seed size of 5\,bp and restricts its evaluation to single-end reads of 100\,bp.}
Other works, such as GenStore~\cite{10.1145/3503222.3507702}, mitigate I/O bottlenecks via in-storage filtering of dissimilar or reference-matching sequences. 
\crviii{None of these pre-alignment filtering approaches leverage both heuristics and hardware acceleration combined with techniques tailored to features of \crviv{\emph{paired-end}} reads.}

\noindent\textbf{GPU-based} solutions~\cite{10568393,bingol2021gatekeeper} leverage parallel processing to accelerate filtering of large genomic datasets. 
Numerous works have studied GPU-based genome analysis~\cite{Sadasivan2023-qx, Dong2024.03.23.586366, marco2021fast, 5161066, gu2023gendp, 10.1145/3698587.3701366}, and some have successfully implemented the \emph{full} read mapping pipeline on GPUs~\cite{gu2023gendp}. As shown in \CR{\S}\ref{sec:e_2_e}, while some GPU implementations achieve higher throughput than CPU solutions, the increased throughput per unit power indicates higher operational costs.
As shown in \CR{\S}\ref{sec:nmsl_eval}, GPUs can efficiently accelerate specific stages of \codealg, such as the SeedMap Query and Light Alignment stages. However, the performance and power efficiency are lower than the \codename{} implementation. Additionally, integrating the GPU into the end-to-end pipeline requires extra orchestration of kernels and data movement to coordinate.

\mypar{DP accelerators} Several works target the compute-intensive dynamic programming (DP) steps such as chaining and alignment. Race logic~\cite{madhavan2014race} encodes timing delays to perform \crviii{Smith-Waterman alignment\cite{SMITH1981195}.} GenDP~\cite{gu2023gendp} introduces a flexible hardware framework for diverse DP patterns. ISA extensions for edge SoCs~\cite{10.1145/3613424.3614306}, vector acceleration frameworks like QUETZAL~\cite{10609714}, and tiling-based approaches like TALCO~\cite{10476438} further aim to optimize DP. GPU-based tools also accelerate chaining~\cite{Sadasivan2023-qx, Dong2024.03.23.586366} and alignment~\cite{marco2021fast, 5161066}. In contrast, \codename{} aims to eliminate reliance on DP, which remains costly in both time and energy even with specialized hardware.

\section{Conclusions}

\crviv{This work presents the first hardware-algorithm co-designed system designed for \crviv{\emph{paired-end}} read mapping. We leverage key insights from an extensive characterization of paired-end read mapping and present a novel read mapping pipeline, \codealg. \codealg introduces a new hash-based filter and a new lightweight alignment
algorithm to increase filtering ratio and reduce computational requirements.
The algorithmic contributions of \codealg are integrated into a hardware-algorithm co-designed accelerator that mitigates the memory access overheads through a memory-access optimized hash-table and benefits from the high-paralellism available within High Bandwidth Memory (HBM). \codename{} delivers end-to-end throughput per unit power improvements of 1575$\times$ and 1.43$\times$ compared to state-of-the-art CPU- and accelerator-based read mappers, respectively. 
It enhances throughput per area by 958$\times$ and 2.35$\times$ over these solutions, while maintaining alignment accuracy comparable to the best available software tools for mapping to the human reference genome.}

\section*{Acknowledgments}
\crviii{We thank the anonymous reviewers of HPCA 2026, MICRO 2025 and ISCA 2025 for their feedback.  SAFARI Research Group thanks the following for research support: the Huawei ZRC Storage Team, the ETH Future Computing Laboratory, AI Chip Center for Emerging Smart Systems (ACCESS), sponsored by InnoHK funding, Hong Kong SAR, and the European Union’s Horizon programme for research and innovation [101047160 - BioPIM].}


\bibliographystyle{unsrt}
\bibliography{references}

\begin{thebibliography}{100}

\bibitem{Kingsmore2024}
Stephen~F. Kingsmore, Russell Nofsinger, and Kasia Ellsworth.
\newblock {Rapid Genomic Sequencing for Genetic Disease Diagnosis and Therapy in Intensive Care Units: A Review}.
\newblock {\em npj Genomic Medicine}, 2024.

\bibitem{cells13060504}
Petar Brlek, Luka Bulić, Matea Bračić, Petar Projić, Vedrana Škaro, Nidhi Shah, Parth Shah, and Dragan Primorac.
\newblock {Implementing Whole Genome Sequencing (WGS) in Clinical Practice: Advantages, Challenges, and Future Perspectives}.
\newblock {\em Cells}, 2024.

\bibitem{clark2019diagnosis}
Michelle~M. Clark, Amber Hildreth, Sergey Batalov, Yan Ding, Shimul Chowdhury, Kelly Watkins, Katarzyna Ellsworth, Brandon Camp, Cyrielle~I. Kint, Calum Yacoubian, Lauge Farnaes, Matthew~N. Bainbridge, Curtis Beebe, Joshua J.~A. Braun, Margaret Bray, Jeanne Carroll, Julie~A. Cakici, Sara~A. Caylor, Christina Clarke, Mitchell~P. Creed, Jennifer Friedman, Alison Frith, Richard Gain, Mary Gaughran, Shauna George, Sheldon Gilmer, Joseph Gleeson, Jeremy Gore, Haiying Grunenwald, Raymond~L. Hovey, Marie~L. Janes, Kejia Lin, Paul~D. McDonagh, Kyle McBride, Patrick Mulrooney, Shareef Nahas, Daeheon Oh, Albert Oriol, Laura Puckett, Zia Rady, Martin~G. Reese, Julie Ryu, Lisa Salz, Erica Sanford, Lawrence Stewart, Nathaly Sweeney, Mari Tokita, Luca Van~Der Kraan, Sarah White, Kristen Wigby, Brett Williams, Terence Wong, Meredith~S. Wright, Catherine Yamada, Peter Schols, John Reynders, Kevin Hall, David Dimmock, Narayanan Veeraraghavan, Thomas Defay, and Stephen~F. Kingsmore.
\newblock {Diagnosis of Genetic Diseases in Seriously Ill Children by Rapid Whole-genome Sequencing and Automated Phenotyping and Interpretation}.
\newblock {\em Science Translational Medicine}, 2019.

\bibitem{farnaes2018rapid}
Lauge Farnaes, Amber Hildreth, Nathaly~M. Sweeney, Michelle~M. Clark, Shimul Chowdhury, Shareef Nahas, Julie~A. Cakici, Wendy Benson, Robert~H. Kaplan, Richard Kronick, Matthew~N. Bainbridge, Jennifer Friedman, Jeffrey~J. Gold, Yan Ding, Narayanan Veeraraghavan, David Dimmock, and Stephen~F. Kingsmore.
\newblock {Rapid Whole-genome Sequencing Decreases Infant Morbidity and Cost of Hospitalization}.
\newblock {\em NPJ Genomic Medicine}, 2018.

\bibitem{sweeney2021rapid}
Nathaly~M. Sweeney, Shareef~A. Nahas, Shimul Chowdhury, Sergey Batalov, Michelle Clark, Sara Caylor, Julie Cakici, John~J. Nigro, Yan Ding, Narayanan Veeraraghavan, Charlotte Hobbs, David Dimmock, and Stephen~F. Kingsmore.
\newblock {Rapid Whole Genome Sequencing Impacts Care and Resource Utilization in Infants with Congenital Heart Disease}.
\newblock {\em NPJ Genomic Medicine}, 2021.

\bibitem{alkan2009personalized}
Can Alkan, Jeffrey~M Kidd, Tomas Marques-Bonet, Gozde Aksay, Francesca Antonacci, Fereydoun Hormozdiari, Jacob~O Kitzman, Carl Baker, Maika Malig, Onur Mutlu, S~Cenk Sahinalp, Richard~A Gibbs, and Evan~E Eichler.
\newblock {Personalized Copy Number and Segmental Duplication Maps Using Next-Generation Sequencing}.
\newblock {\em Nature Genetics}, 2009.

\bibitem{ginsburg2009genomic}
Geoffrey~S Ginsburg and Huntington~F Willard.
\newblock {Genomic and Personalized Medicine: Foundations and Applications}.
\newblock {\em Translational Research}, 2009.

\bibitem{chin2011cancer}
Lynda Chin, Jannik~N Andersen, and P~Andrew Futreal.
\newblock {Cancer Genomics: From Discovery Science to Personalized Medicine}.
\newblock {\em Nature Medicine}, 2011.

\bibitem{Bandoy2021}
D.~J. Darwin~R. Bandoy and Bart~C. Weimer.
\newblock {Analysis of SARS-CoV-2 Genomic Epidemiology reveals Disease Transmission Coupled to Variant Emergence and Allelic Variation}.
\newblock {\em Scientific Reports}, 2021.

\bibitem{ZHENG2024121513}
Xiawan Zheng, Keyue Zhao, Bingjie Xue, Yu~Deng, Xiaoqing Xu, Weifu Yan, Chao Rong, Kathy Leung, Joseph~T. Wu, Gabriel~M. Leung, Malik Peiris, Leo~L.M. Poon, and Tong Zhang.
\newblock Tracking diarrhea viruses and mpox virus using the wastewater surveillance network in hong kong.
\newblock {\em Water Research}, 2024.

\bibitem{bloom2021massively}
Joshua~S. Bloom, Laila Sathe, Chetan Munugala, Eric~M. Jones, Molly Gasperini, Nathan~B. Lubock, Fauna Yarza, Erin~M. Thompson, Kyle~M. Kovary, Jimin Park, Dawn Marquette, Stephania Kay, Mark Lucas, TreQuan Love, A.~Sina~Booeshaghi, Oliver~F. Brandenberg, Longhua Guo, James Boocock, Myles Hochman, Scott~W. Simpkins, Isabella Lin, Nathan LaPierre, Duke Hong, Yi~Zhang, Gabriel Oland, Bianca~Judy Choe, Sukantha Chandrasekaran, Evann~E. Hilt, Manish~J. Butte, Robert Damoiseaux, Clifford Kravit, Aaron~R. Cooper, Yi~Yin, Lior Pachter, Omai~B. Garner, Jonathan Flint, Eleazar Eskin, Chongyuan Luo, Sriram Kosuri, Leonid Kruglyak, and Valerie~A. Arboleda.
\newblock {Massively Scaled-up Testing for SARS-CoV-2 RNA via Next-generation Sequencing of Pooled and Barcoded Nasal and Saliva Samples}.
\newblock {\em Nature Biomedical Engineering}, 2021.

\bibitem{yelagandula2021multiplexed}
Ramesh Yelagandula, Aleksandr Bykov, Alexander Vogt, Robert Heinen, Ezgi {\"O}zkan, Marcus~Martin Strobl, Juliane~Christina Baar, Kristina Uzunova, Bence Hajdusits, Darja Kordic, Erna Suljic, Amina Kurtovic-Kozaric, Sebija Izetbegovic, Justine Schaeffer, Peter Hufnagl, Alexander Zoufaly, Tamara Seitz, Mariam Al-Rawi, Stefan Ameres, Juliane Baar, Benedikt Bauer, Nikolaus Beer, Katharina Bergauer, Wolfgang Binder, Claudia Blaukopf, Boril Bochev, Julius Brennecke, Selina Brinnich, Aleksandra Bundalo, Meinrad Busslinger, Tim Clausen, Geert de~Vries, Marcus Dekens, David Drechsel, Zuzana Dzupinkova, Michaela Eckmann-Mader, Michaela Fellner, Thomas Fellner, Laura Fin, Bianca~Valeria Gapp, Gerlinde Grabmann, Irina Grishkovskaya, Astrid Hagelkruys, Dominik Handler, David Haselbach, Louisa Hempel, Louisa Hill, David Hoffmann, Stefanie Horer, Harald Isemann, Robert Kalis, Max Kellner, Juliane Kley, Thomas K{\"o}cher, Alwin K{\"o}hler, Christian Krauditsch, Sabina Kula, Sonja Lang, Richard Latham, Marie-Christin
  Leitner, Thomas Leonard, Dominik Lindenhofer, Raphael~Arthur Manzenreither, Martin Matl, Karl Mechtler, Anton Meinhart, Stefan Mereiter, Thomas Micheler, Paul Moeseneder, Tobias Neumann, Simon Nimpf, Magnus Nordborg, Egon Ogris, Michaela Pagani, Andrea Pauli, Jan-Michael Peters, Petra Pjevac, Clemens Plaschka, Martina Rath, Daniel Reumann, Sarah Rieser, Marianne Rocha-Hasler, Alan Rodriguez, Nathalie Ropek, James~Julian Ross, Harald Scheuch, Karina Schindler, Clara Schmidt, Hannes Schmidt, Jakob Schnabl, Stefan Sch{\"u}chner, Tanja Schwickert, Andreas Sommer, Daniele Soldoroni, Johannes Stadlmann, Peter Steinlein, Marcus Strobl, Simon Strobl, Qiong Sun, Wen Tang, Linda Tr{\"u}bestein, Johanna Trupke, Christian Umkehrer, Sandor Urmosi-Incze, Gijs Versteeg, Vivien Vogt, Michael Wagner, Martina Weissenboeck, Barbara Werner, Johannes Zuber, Manuela F{\"o}dinger, Franz Allerberger, Alexander Stark, Luisa Cochella, Ulrich Elling, and {VCDI}.
\newblock {Multiplexed Detection of SARS-CoV-2 and Other Respiratory Infections in High Throughput by SARSeq}.
\newblock {\em Nature Communications}, 2021.

\bibitem{le2013selected}
Vien Thi~Minh Le and Binh~An Diep.
\newblock {Selected Insights from Application of Whole Genome Sequencing for Outbreak Investigations}.
\newblock {\em Current Opinion in Critical Care}, 2013.

\bibitem{nikolayevskyy2016whole}
Vlad Nikolayevskyy, Katharina Kranzer, Stefan Niemann, and Francis Drobniewski.
\newblock {Whole Genome Sequencing of Mycobacterium Tuberculosis for Detection of Recent Transmission and Tracing Outbreaks: A Systematic Review}.
\newblock {\em Tuberculosis}, 2016.

\bibitem{qiu2015whole}
Shaofu Qiu, Peng Li, Hongbo Liu, Yong Wang, Nan Liu, Chengyi Li, Shenlong Li, Ming Li, Zhengjie Jiang, Huandong Sun, Ying Li, Jing Xie, Chaojie Yang, Jian Wang, Hao Li, Shengjie Yi, Zhihao Wu, Leili Jia, Ligui Wang, Rongzhang Hao, Yansong Sun, Liuyu Huang, Hui Ma, Zhengquan Yuan, and Hongbin Song.
\newblock {Whole-genome Sequencing for Tracing the Transmission Link between Two ARD Outbreaks Caused by A Novel HAdV Serotype 7 Variant, China}.
\newblock {\em Scientific Reports}, 2015.

\bibitem{gilchrist2015whole}
Carol~A Gilchrist, Stephen~D Turner, Margaret~F Riley, William~A Petri, and Erik~L Hewlett.
\newblock {Whole-genome Sequencing in Outbreak Analysis}.
\newblock {\em Clinical Microbiology Reviews}, 2015.

\bibitem{prasad2021soil}
Shiv Prasad, Lal~Chand Malav, Jairam Choudhary, Sudha Kannojiya, Monika Kundu, Sandeep Kumar, and Ajar~Nath Yadav.
\newblock {\em Soil Microbiomes for Healthy Nutrient Recycling}.
\newblock 2021.

\bibitem{Mascher2024}
Martin Mascher, Murukarthick Jayakodi, Hyeonah Shim, and Nils Stein.
\newblock Promises and challenges of crop translational genomics.
\newblock {\em Nature}, 2024.

\bibitem{Schreiber2024}
Mona Schreiber, Murukarthick Jayakodi, Nils Stein, and Martin Mascher.
\newblock Plant pangenomes for crop improvement, biodiversity and evolution.
\newblock {\em Nature Reviews Genetics}, 2024.

\bibitem{urbanek2018degradation}
Aneta~K. Urbanek, Waldemar Rymowicz, and Aleksandra~M. Miro{\'{n}}czuk.
\newblock Degradation of plastics and plastic-degrading bacteria in cold marine habitats.
\newblock {\em Applied Microbiology and Biotechnology}, 2018.

\bibitem{edgar2022petabase}
Robert~C. Edgar, Jeff Taylor, Victor Lin, Tomer Altman, Pierre Barbera, Dmitry Meleshko, Dan Lohr, Gherman Novakovsky, Benjamin Buchfink, Basem Al-Shayeb, Jillian~F. Banfield, Marcos de~la Pe{\~{n}}a, Anton Korobeynikov, Rayan Chikhi, and Artem Babaian.
\newblock Petabase-scale sequence alignment catalyses viral discovery.
\newblock {\em Nature}, 2022.

\bibitem{paoli2022biosynthetic}
Lucas Paoli, Hans-Joachim Ruscheweyh, Clarissa~C. Forneris, Florian Hubrich, Satria Kautsar, Agneya Bhushan, Alessandro Lotti, Quentin Clayssen, Guillem Salazar, Alessio Milanese, Charlotte~I. Carlstr{\"o}m, Chrysa Papadopoulou, Daniel Gehrig, Mikhail Karasikov, Harun Mustafa, Martin Larralde, Laura~M. Carroll, Pablo S{\'a}nchez, Ahmed~A. Zayed, Dylan~R. Cronin, Silvia~G. Acinas, Peer Bork, Chris Bowler, Tom~O. Delmont, Josep~M. Gasol, Alvar~D. Gossert, Andr{\'e} Kahles, Matthew~B. Sullivan, Patrick Wincker, Georg Zeller, Serina~L. Robinson, J{\"o}rn Piel, and Shinichi Sunagawa.
\newblock Biosynthetic potential of the global ocean microbiome.
\newblock {\em Nature}, 2022.

\bibitem{02233705-202312000-00003}
Chunsheng Zhao, Ziwei Zhang, Linlin Sun, Ronglu Bai, Lizhi Wang, and Shilin Chen.
\newblock {Genome Sequencing provides Potential Strategies for Drug Discovery and Synthesis}.
\newblock {\em Acupuncture and Herbal Medicine}, 2023.

\bibitem{doi:10.1128/spectrum.03617-22}
Murari Bhandari, Irani~U. Rathnayake, Flavia Huygens, Son Nguyen, Brett Heron, and Amy~V. Jennison.
\newblock {Genomic and Evolutionary Insights into Australian Toxigenic Vibrio cholerae O1 Strains}.
\newblock {\em Microbiology Spectrum}, 2023.

\bibitem{Chiu2023}
Kuo-Ping Chiu, Lutimba Stuart, Hong~Sain Ooi, John Yu, David~Glenn Smith, and Kurtis Jai-Chyi Pei.
\newblock {Genome Sequencing and Application of Taiwanese Macaque Macaca Cyclopis}.
\newblock {\em Scientific Reports}, 2023.

\bibitem{shi2018evolutionary}
Mang Shi, Xian-Dan Lin, Xiao Chen, Jun-Hua Tian, Liang-Jun Chen, Kun Li, Wen Wang, John-Sebastian Eden, Jin-Jin Shen, Li~Liu, et~al.
\newblock {The Evolutionary History of Vertebrate RNA Viruses}.
\newblock {\em Nature}, 2018.

\bibitem{werner2012comparison}
Jeffrey~J Werner, Dennis Zhou, J~Gregory Caporaso, Rob Knight, and Largus~T Angenent.
\newblock {Comparison of Illumina paired-end and single-direction Sequencing for Microbial 16S rRNA Gene Amplicon Surveys}.
\newblock {\em The ISME journal}, 2012.

\bibitem{bartram2011generation}
Andrea~K Bartram, Michael~DJ Lynch, Jennifer~C Stearns, Gabriel Moreno-Hagelsieb, and Josh~D Neufeld.
\newblock {Generation of Multimillion-sequence 16S rRNA Gene Libraries from Complex Microbial Communities by Assembling paired-end Illumina reads}.
\newblock {\em {Applied and Environmental Microbiology}}, 2011.

\bibitem{campbell2008identification}
Peter~J Campbell, Philip~J Stephens, Erin~D Pleasance, Sarah O'Meara, Heng Li, Thomas Santarius, Lucy~A Stebbings, Catherine Leroy, Sarah Edkins, Claire Hardy, et~al.
\newblock {Identification of Somatically Acquired Rearrangements in Cancer using Genome-wide Massively Parallel Paired-end Sequencing}.
\newblock {\em Nature genetics}, 2008.

\bibitem{borsting2015next}
Claus B{\o}rsting and Niels Morling.
\newblock {Next Generation Sequencing and its Applications in Forensic Genetics}.
\newblock {\em Forensic Science International: Genetics}, 2015.

\bibitem{rizzo2012key}
Jason~M Rizzo and Michael~J Buck.
\newblock {Key Principles and Clinical Applications of “Next-Generation” DNA Sequencing}.
\newblock {\em Cancer prevention research}, 2012.

\bibitem{edgren2011identification}
Henrik Edgren, Astrid Murumagi, Sara Kangaspeska, Daniel Nicorici, Vesa Hongisto, Kristine Kleivi, Inga~H Rye, Sandra Nyberg, Maija Wolf, Anne-Lise Borresen-Dale, et~al.
\newblock {Identification of Fusion Genes in Breast Cancer by paired-end RNA-sequencing}.
\newblock {\em Genome biology}, 2011.

\bibitem{illumina}
Illumina.
\newblock {NovaSeq 6000 System Specifications}.
\newblock \url{https://emea.illumina.com/systems/sequencing-platforms/novaseq/specifications.html}, 2020.

\bibitem{PacBio2025}
Inc. Pacific Biosciences~of California.
\newblock Pacbio.
\newblock \url{https://www.pacb.com/}, 2025.

\bibitem{NanoporeTech2025}
Nanopore technologies.
\newblock \url{https://nanoporetech.com/}, 2025.

\bibitem{illumina_paired_end_2025}
Illumina.
\newblock {Paired-end vs single-read sequencing}.
\newblock \url{https://emea.illumina.com/science/technology/next-generation-sequencing/plan-experiments/paired-end-vs-single-read.html}.

\bibitem{https://doi.org/10.1002/0471250953.bi1506s45}
Xian Fan, Travis~E. Abbott, David Larson, and Ken Chen.
\newblock {BreakDancer: Identification of Genomic Structural Variation from Paired-End Read Mapping}.
\newblock {\em Current Protocols in Bioinformatics}, 2014.

\bibitem{Freedman2020}
Adam~H. Freedman, John~M. Gaspar, and Timothy~B. Sackton.
\newblock {Short Paired-end Reads trump Long Single-end Reads for Expression Analysis}.
\newblock {\em BMC Bioinformatics}, 2020.

\bibitem{Zhu2015}
Xiao Zhu, Henry C.~M. Leung, Rongjie Wang, Francis Y.~L. Chin, Siu~Ming Yiu, Guangri Quan, Yajie Li, Rui Zhang, Qinghua Jiang, Bo~Liu, Yucui Dong, Guohui Zhou, and Yadong Wang.
\newblock {misFinder: Identify mis-assemblies in an Unbiased Manner Using Reference and Paired-end Reads}.
\newblock {\em BMC Bioinformatics}, 2015.

\bibitem{10.1093/bioinformatics/btq152}
Iman Hajirasouliha, Fereydoun Hormozdiari, Can Alkan, Jeffrey~M. Kidd, Inanc Birol, Evan~E. Eichler, and S.~Cenk Sahinalp.
\newblock {Detection and Characterization of Novel Sequence Insertions Using Paired-end Next-Generation Sequencing}.
\newblock {\em Bioinformatics}, 2010.

\bibitem{eren2013filtering}
A~Murat Eren, Joseph~H Vineis, Hilary~G Morrison, and Mitchell~L Sogin.
\newblock {A Filtering Method to generate High Quality Short Reads using Illumina Paired-end Technology}.
\newblock {\em PloS one}, 2013.

\bibitem{grimm2013accurate}
Dominik Grimm, J{\"o}rg Hagmann, Daniel Koenig, Detlef Weigel, and Karsten Borgwardt.
\newblock {Accurate Indel Prediction using Paired-end Short Reads}.
\newblock {\em BMC genomics}, 2013.

\bibitem{cameron2019comprehensive}
Daniel~L Cameron, Leon Di~Stefano, and Anthony~T Papenfuss.
\newblock {Comprehensive Evaluation and Characterisation of Short Read General-purpose Structural Variant Calling Software}.
\newblock {\em Nature communications}, 2019.

\bibitem{rausch2012delly}
Tobias Rausch, Thomas Zichner, Andreas Schlattl, Adrian~M St{\"u}tz, Vladimir Benes, and Jan~O Korbel.
\newblock {DELLY: Structural Variant Discovery by Integrated Paired-end and Split-read Analysis}.
\newblock {\em Bioinformatics}, 2012.

\bibitem{mutlu2023accelerating}
Onur Mutlu and Can Firtina.
\newblock {Accelerating Genome Analysis via Algorithm-Architecture Co-Design}.
\newblock In {\em DAC}, 2023.

\bibitem{alser2022molecules}
Mohammed Alser, Joel Lindegger, Can Firtina, Nour Almadhoun, Haiyu Mao, Gagandeep Singh, Juan Gomez-Luna, and Onur Mutlu.
\newblock From molecules to genomic variations: Accelerating genome analysis via intelligent algorithms and architectures.
\newblock {\em Computational and Structural Biotechnology Journal}, 2022.

\bibitem{alser2020accelerating}
Mohammed Alser, Z{\"u}lal Bing{\"o}l, Damla~Senol Cali, Jeremie Kim, Saugata Ghose, Can Alkan, and Onur Mutlu.
\newblock {Accelerating Genome Analysis: A Primer on an Ongoing Journey}.
\newblock {\em IEEE Micro}, 2020.

\bibitem{10.1145/3503222.3507702}
Nika Mansouri~Ghiasi, Jisung Park, Harun Mustafa, Jeremie Kim, Ataberk Olgun, Arvid Gollwitzer, Damla Senol~Cali, Can Firtina, Haiyu Mao, Nour Almadhoun~Alserr, Rachata Ausavarungnirun, Nandita Vijaykumar, Mohammed Alser, and Onur Mutlu.
\newblock {GenStore: a high-performance in-storage processing system for genome sequence analysis}.
\newblock ASPLOS, 2022.

\bibitem{alser2020sneakysnake}
Mohammed Alser, Taha Shahroodi, Juan G{\'o}mez-Luna, Can Alkan, and Onur Mutlu.
\newblock {SneakySnake: A Fast and Accurate Universal Genome Pre-alignment Filter for CPUs, GPUs and FPGAs}.
\newblock {\em Bioinformatics}, 2020.

\bibitem{alser2017gatekeeper}
Mohammed Alser, Hasan Hassan, Hongyi Xin, O{\u{g}}uz Ergin, Onur Mutlu, and Can Alkan.
\newblock {GateKeeper: A New Hardware Architecture for Accelerating Pre-alignment in DNA Short Read Mapping}.
\newblock {\em Bioinformatics}, 2017.

\bibitem{xin2013accelerating}
Hongyi Xin, Donghyuk Lee, Farhad Hormozdiari, Samihan Yedkar, Onur Mutlu, and Can Alkan.
\newblock {Accelerating Read Mapping with FastHASH}.
\newblock {\em BMC Genomics}, 2013.

\bibitem{xin2015shifted}
Hongyi Xin, John Greth, John Emmons, Gennady Pekhimenko, Carl Kingsford, Can Alkan, and Onur Mutlu.
\newblock {Shifted Hamming Distance: A Fast and Accurate SIMD-friendly Filter to Accelerate Alignment Verification in Read Mapping}.
\newblock {\em Bioinformatics}, 2015.

\bibitem{kim2018grim}
Jeremie~S Kim, Damla~Senol Cali, Hongyi Xin, Donghyuk Lee, Saugata Ghose, Mohammed Alser, Hasan Hassan, Oguz Ergin, Can Alkan, and Onur Mutlu.
\newblock {GRIM-Filter: Fast Seed Location Filtering in DNA Read Mapping Using Processing-in-memory Technologies}.
\newblock {\em BMC Genomics}, 2018.

\bibitem{alser2019shouji}
Mohammed Alser, Hasan Hassan, Akash Kumar, Onur Mutlu, and Can Alkan.
\newblock {Shouji: A Fast and Efficient Pre-alignment Filter for Sequence Alignment}.
\newblock {\em Bioinformatics}, 2019.

\bibitem{Liyanage2023}
Kisaru Liyanage, Hiruna Samarakoon, Sri Parameswaran, and Hasindu Gamaarachchi.
\newblock {Efficient End-to-End Long-read Sequence Mapping using minimap2-fpga integrated with Hardware Accelerated Chaining}.
\newblock {\em Scientific Reports}, 2023.

\bibitem{6239809}
Corey~B. Olson, Maria Kim, Cooper Clauson, Boris Kogon, Carl Ebeling, Scott Hauck, and Walter~L. Ruzzo.
\newblock {Hardware Acceleration of Short Read Mapping}.
\newblock In {\em FCCM}, 2012.

\bibitem{Samarasinghe2021Energy}
S.~Samarasinghe, P.~Premathilaka, W.~Herath, H.~Gamaarachchi, and R.~Ragel.
\newblock {Energy Efficient Adaptive Banded Event Alignment using OpenCL on FPGAs}.
\newblock In {\em ICIIS}, 2021.

\bibitem{koliogeorgi2022gandafl}
Konstantina Koliogeorgi, Sotirios Xydis, Georgi Gaydadjiev, and Dimitrios Soudris.
\newblock {Gandafl: Dataflow Acceleration for Short Read Alignment on NGS Data}.
\newblock {\em IEEE TC}, 2022.

\bibitem{guo2019hardware}
Licheng Guo, Jason Lau, Zhenyuan Ruan, Peng Wei, and Jason Cong.
\newblock {Hardware Acceleration of Long Read Pairwise Overlapping in Genome Sequencing: A Race between FPGA and GPU}.
\newblock In {\em FCCM}, 2019.

\bibitem{singh2021fpga}
Gagandeep Singh, Mohammed Alser, Damla~Senol Cali, Dionysios Diamantopoulos, Juan G{\'o}mez-Luna, Henk Corporaal, and Onur Mutlu.
\newblock {FPGA-based Near-Memory Acceleration of Modern Data-Intensive Applications}.
\newblock {\em IEEE Micro}, 2021.

\bibitem{chen2021high}
Yen-Lung Chen, Bo-Yi Chang, Chia-Hsiang Yang, and Tzi-Dar Chiueh.
\newblock {A High-Throughput FPGA Accelerator for Short-Read Mapping of the Whole Human Genome}.
\newblock {\em IEEE TPDS}, 2021.

\bibitem{fei2018fpgasw}
Xia Fei, Zou Dan, Lu~Lina, Man Xin, and Zhang Chunlei.
\newblock {FPGASW: Accelerating Large-scale Smith--Waterman Sequence Alignment Application with Backtracking on FPGA Linear Systolic Array}.
\newblock {\em Interdisciplinary Sciences: Computational Life Sciences}, 2018.

\bibitem{haghi2021fpga}
Abbas Haghi, Santiago Marco-Sola, Lluc Alvarez, Dionysios Diamantopoulos, Christoph Hagleitner, and Miquel Moreto.
\newblock {An FPGA Accelerator of the Wavefront Algorithm for Genomics Pairwise Alignment}.
\newblock In {\em FPL}, 2021.

\bibitem{rucci2018swifold}
Enzo Rucci, Carlos Garcia, Guillermo Botella, Armando De~Giusti, Marcelo Naiouf, and Manuel Prieto-Matias.
\newblock {SWIFOLD: Smith-Waterman Implementation on FPGA with OpenCL for Long DNA Sequences}.
\newblock {\em BMC Systems Biology}, 2018.

\bibitem{li2021pipebsw}
Luyi Li, Jun Lin, and Zhongfeng Wang.
\newblock {PipeBSW: A Two-Stage Pipeline Structure for Banded Smith-Waterman Algorithm on FPGA}.
\newblock In {\em ISVLSI}, 2021.

\bibitem{liao2018adaptively}
Yi-Lun Liao, Yu-Cheng Li, Nae-Chyun Chen, and Yi-Chang Lu.
\newblock {Adaptively Banded Smith-Waterman Algorithm for Long Reads and its Hardware Accelerator}.
\newblock In {\em ASAP}, 2018.

\bibitem{fujiki2018genax}
Daichi Fujiki, Arun Subramaniyan, Tianjun Zhang, Yu~Zeng, Reetuparna Das, David Blaauw, and Satish Narayanasamy.
\newblock {Genax: A Genome Sequencing Accelerator}.
\newblock In {\em ISCA}, 2018.

\bibitem{nag2019gencache}
Anirban Nag, CN~Ramachandra, Rajeev Balasubramonian, Ryan Stutsman, Edouard Giacomin, Hari Kambalasubramanyam, and Pierre-Emmanuel Gaillardon.
\newblock {GenCache: Leveraging In-cache Operators for Efficient Sequence Alignment}.
\newblock In {\em MICRO}, 2019.

\bibitem{turakhia2018darwin}
Yatish Turakhia, Gill Bejerano, and William~J Dally.
\newblock {Darwin: A Genomics Co-processor Provides up to 15,000 x Acceleration on Long Read Assembly}.
\newblock In {\em ASPLOS}, 2018.

\bibitem{Lindegger2023Scrooge}
J.~Lindegger, D.~Senol~Cali, M.~Alser, J.~Gómez-Luna, N.~M. Ghiasi, and O.~Mutlu.
\newblock {Scrooge: a Fast and Memory-Frugal Genomic Sequence Aligner for CPUs, GPUs, and ASICs}.
\newblock {\em Bioinformatics}, 2023.

\bibitem{cali2020genasm}
Damla~Senol Cali, Gurpreet~S. Kalsi, Zülal Bingöl, Can Firtina, Lavanya Subramanian, Jeremie~S. Kim, Rachata Ausavarungnirun, Mohammed Alser, Juan Gomez-Luna, Amirali Boroumand, Anant Norion, Allison Scibisz, Sreenivas Subramoneyon, Can Alkan, Saugata Ghose, and Onur Mutlu.
\newblock {GenASM: A High-Performance, Low-Power Approximate String Matching Acceleration Framework for Genome Sequence Analysis}.
\newblock In {\em MICRO}, 2020.

\bibitem{banerjee2018asap}
Subho~Sankar Banerjee, Mohamed El-Hadedy, Jong~Bin Lim, Zbigniew~T Kalbarczyk, Deming Chen, Steven~S Lumetta, and Ravishankar~K Iyer.
\newblock {ASAP: Accelerated Short-read Alignment on Programmable Hardware}.
\newblock {\em IEEE Transactions on Computers}, 2019.

\bibitem{fujiki2020seedex}
Daichi Fujiki, Shunhao Wu, Nathan Ozog, Kush Goliya, David Blaauw, Satish Narayanasamy, and Reetuparna Das.
\newblock {SeedEx: A Genome Sequencing Accelerator for Optimal Alignments in Subminimal Space}.
\newblock In {\em MICRO}, 2020.

\bibitem{cali2022segram}
Damla~Senol Cali, Konstantinos Kanellopoulos, Jo{\"e}l Lindegger, Z{\"u}lal Bing{\"o}l, Gurpreet~S Kalsi, Ziyi Zuo, Can Firtina, Meryem~Banu Cavlak, Jeremie Kim, Nika~Mansouri Ghiasi, et~al.
\newblock Segram: A universal hardware accelerator for genomic sequence-to-graph and sequence-to-sequence mapping.
\newblock In {\em ISCA}, 2022.

\bibitem{liu2010cudasw++}
Yongchao Liu, Bertil Schmidt, and Douglas~L Maskell.
\newblock {CUDASW++ 2.0: Enhanced Smith-Waterman Protein Database Search on CUDA-enabled GPUs Based on SIMT and Virtualized SIMD Abstractions}.
\newblock {\em BMC Research Notes}, 2010.

\bibitem{liu2013cudasw++}
Yongchao Liu, Adrianto Wirawan, and Bertil Schmidt.
\newblock {CUDASW++ 3.0: Accelerating Smith-Waterman Protein Database Search by Coupling CPU and GPU SIMD Instructions}.
\newblock {\em BMC Bioinformatics}, 2013.

\bibitem{nishimura2017accelerating}
Takahiro Nishimura, Jacir~L Bordim, Yasuaki Ito, and Koji Nakano.
\newblock {Accelerating the Smith-waterman Algorithm Using Bitwise Parallel Bulk Computation Technique on GPU}.
\newblock In {\em IPDPSW}, 2017.

\bibitem{gu2023gendp}
Yufeng Gu, Arun Subramaniyan, Tim Dunn, Alireza Khadem, Kuan-Yu Chen, Somnath Paul, Md~Vasimuddin, Sanchit Misra, David Blaauw, Satish Narayanasamy, and Reetuparna Das.
\newblock {GenDP: A Framework of Dynamic Programming Acceleration for Genome Sequencing Analysis}.
\newblock In {\em ISCA}, 2023.

\bibitem{li2018minimap2}
Heng Li.
\newblock {Minimap2: Pairwise Alignment for Nucleotide Sequences}.
\newblock {\em Bioinformatics}, 2018.

\bibitem{houtgast2017efficient}
Ernst~Joachim Houtgast, VladMihai Sima, Koen Bertels, and Zaid AlArs.
\newblock {An Efficient GPU-accelerated Implementation of Genomic Short Read Mapping with BWA-MEM}.
\newblock {\em ACM SIGARCH Computer Architecture News}, 2017.

\bibitem{baichoo2017computational}
Shakuntala Baichoo and Christos~A Ouzounis.
\newblock {Computational Complexity of Algorithms for Sequence Comparison, Short-Read Assembly and Genome Alignment}.
\newblock {\em Biosystems}, 2017.

\bibitem{SMITH1981195}
T.F. Smith and M.S. Waterman.
\newblock {Identification of Common Molecular Subsequences}.
\newblock {\em Journal of Molecular Biology}, 1981.

\bibitem{needleman1970general}
Saul~B Needleman and Christian~D Wunsch.
\newblock {A General Method Applicable to the Search for Similarities in the Amino Acid Sequence of Two Proteins}.
\newblock {\em Journal of Molecular Biology}, 1970.

\bibitem{Li2009SAM}
Heng Li, Bob Handsaker, Alec Wysoker, Tim Fennell, Jue Ruan, Nils Homer, Gabor Marth, Goncalo Abecasis, and Richard Durbin.
\newblock {The Sequence Alignment/Map format and SAMtools}.
\newblock {\em Bioinformatics}, 2009.

\bibitem{Alser2025}
Mohammed Alser, Julien Eudine, and Onur Mutlu.
\newblock Taming large-scale genomic analyses via sparsified genomics.
\newblock {\em Nature Communications}, 2025.

\bibitem{mordorintelligence_short_read_sequencing_2025}
Mordor Intelligence.
\newblock {Short-read sequencing market}.
\newblock \url{https://www.mordorintelligence.com/industry-reports/short-read-sequencing-market}.

\bibitem{iontorrent}
Thermo~Fisher Scientific.
\newblock {Ion Torrent}.
\newblock \url{https://www.thermofisher.com/ch/en/home.html}.

\bibitem{giab_hg002}
{Genome in a Bottle}.
\newblock {HG002 (NA24385, Ashkenazim Trio) BGIseq 2x150bp 100x Data}.
\newblock \url{https://ftp.ncbi.nlm.nih.gov/ReferenceSamples/giab/data/AshkenazimTrio/HG002_NA24385_son/NIST_BGIseq_2x150bp_100x/}, 2023.

\bibitem{zook2016extensive}
Justin~M. Zook, David Catoe, Jennifer McDaniel, Lindsay Vang, Noah Spies, Arend Sidow, Ziming Weng, Yuling Liu, Christopher~E. Mason, Noah Alexander, Elizabeth Henaff, Alexa~B.R. McIntyre, Dhruva Chandramohan, Feng Chen, Erich Jaeger, Ali Moshrefi, Khoa Pham, William Stedman, Tiffany Liang, Michael Saghbini, Zeljko Dzakula, Alex Hastie, Han Cao, Gintaras Deikus, Eric Schadt, Robert Sebra, Ali Bashir, Rebecca~M. Truty, Christopher~C. Chang, Natali Gulbahce, Keyan Zhao, Srinka Ghosh, Fiona Hyland, Yutao Fu, Mark Chaisson, Chunlin Xiao, Jonathan Trow, Stephen~T. Sherry, Alexander~W. Zaranek, Madeleine Ball, Jason Bobe, Preston Estep, George~M. Church, Patrick Marks, Sofia Kyriazopoulou-Panagiotopoulou, Grace~X.Y. Zheng, Michael Schnall-Levin, Heather~S. Ordonez, Patrice~A. Mudivarti, Kristina Giorda, Ying Sheng, Karoline~Bjarnesdatter Rypdal, and Marc Salit.
\newblock {Extensive Sequencing of Seven Human Genomes to Characterize Benchmark Reference Materials}.
\newblock {\em Scientific data}, 2016.

\bibitem{giab_grch38}
{Genome in a Bottle Consortium}.
\newblock {GRCh38 Reference Genome (GCA\_000001405.15\_GRCh38\_no\_alt\_analysis\_set.fasta.gz)}.
\newblock \url{https://ftp.ncbi.nlm.nih.gov/ReferenceSamples/giab/release/references/GRCh38/GCA_000001405.15_GRCh38_no_alt_analysis_set.fasta.gz}, 2023.

\bibitem{9154510}
Mohammed Alser, Zülal Bingöl, Damla~Senol Cali, Jeremie Kim, Saugata Ghose, Can Alkan, and Onur Mutlu.
\newblock {Accelerating Genome Analysis: A Primer on an Ongoing Journey}.
\newblock {\em IEEE Micro}, 2020.

\bibitem{xxhash2024}
Yann Collet.
\newblock {xxHash - Extremely fast hash algorithm}.
\newblock \url{https://github.com/Cyan4973/xxHash}, 2024.

\bibitem{fasthashsource}
{Hongyi Xin, Donghyuk Lee, Farhad Hormozdiari, Samihan Yedkar, Onur Mutlu and Can Alkan}.
\newblock {FastHASH Source Code}.
\newblock \url{https://github.com/CMU-SAFARI/FastHASH}.

\bibitem{shdsource}
{Hongyi Xin , John Greth , John Emmons , Gennady Pekhimenko , Carl Kingsford , Can Alkan , Onur Mutlu}.
\newblock {Shifted Hamming Distance Source Code}.
\newblock \url{https://github.com/CMU-SAFARI/Shifted-Hamming-Distance}.

\bibitem{10.1145/3240302.3240315}
Shang Li, Dhiraj Reddy, and Bruce Jacob.
\newblock {A performance \& power comparison of modern high-speed DRAM architectures}.
\newblock In {\em MEMSYS}, 2018.

\bibitem{luo2023ramulator2}
Haocong Luo, Yahya~Can Tu\u{g}rul, F.~Nisa Bostancı, Ataberk Olgun, A.~Giray Ya\u{g}l{\i}k\c{c}{\i}, and Onur Mutlu.
\newblock {Ramulator 2.0: A Modern, Modular, and Extensible DRAM Simulator}, 2023.

\bibitem{10.1145/3577193.3593703}
Minh Pham, Yicheng Tu, and Xiaoyi Lv.
\newblock {Accelerating BWA-MEM Read Mapping on GPUs}.
\newblock In {\em ICS}, 2023.

\bibitem{guide2011intel}
Intel Guide.
\newblock Intel{\textregistered} 64 and ia-32 architectures software developer’s manual.
\newblock {\em Volume 3B: System programming Guide, Part}, 2, 2011.

\bibitem{nvprof}
NVIDIA.
\newblock {Profiler User’s Guide}.
\newblock \url{https://docs.nvidia.com/cuda/profiler-users-guide/index.html}, 2024.

\bibitem{ramulatorsource2}
{Haocong Luo, Yahya Can Tugrul, F. Nisa Bostancı, Ataberk Olgun, A. Giray Yaglıkcı, and Onur Mutlu}.
\newblock {Ramulator 2.0 Source Code}.
\newblock \url{https://github.com/CMU-SAFARI/ramulator2}.

\bibitem{ramulatorsource}
Yoongu Kim, Weikun Yang, and Onur Mutlu.
\newblock {Ramulator Source Code}.
\newblock \url{https://github.com/CMU-SAFARI/ramulator}.

\bibitem{kim2016ramulator}
Yoongu Kim, Weikun Yang, and Onur Mutlu.
\newblock {Ramulator: A Fast and Extensible DRAM Simulator}.
\newblock {\em CAL}, 2015.

\bibitem{8999595}
Shang Li, Zhiyuan Yang, Dhiraj Reddy, Ankur Srivastava, and Bruce Jacob.
\newblock {DRAMsim3: A Cycle-Accurate, Thermal-Capable DRAM Simulator}.
\newblock {\em IEEE CAL}, 2020.

\bibitem{hbm2e}
JEDEC Solid State~Technology Association.
\newblock {“JESD235C: High Bandwidth Memory (HBM) DRAM}, 2020.

\bibitem{SystemVerilogIEEE}
{IEEE Standard for SystemVerilog}.
\newblock \url{https://standards.ieee.org/standard/1800-2017.html}, 2018.

\bibitem{iwai1999cmos}
Hiroshi Iwai.
\newblock Cmos technology-year 2010 and beyond.
\newblock {\em IEEE Journal of Solid-State Circuits}, 1999.

\bibitem{cacti}
Hewlett Packard.
\newblock {CACTI: A tool for modeling the power, area, and timing of memory hierarchies}, 2024.

\bibitem{10.1145/3085572}
Rajeev Balasubramonian, Andrew~B. Kahng, Naveen Muralimanohar, Ali Shafiee, and Vaishnav Srinivas.
\newblock {CACTI 7: New Tools for Interconnect Exploration in Innovative Off-Chip Memories}.
\newblock {\em TACO}, 2017.

\bibitem{STILLMAKER201774}
{Scaling equations for the accurate prediction of CMOS device performance from 180nm to 7nm}.
\newblock {\em Integration}, 2017.

\bibitem{garrison2012haplotypebasedvariantdetectionshortread}
Erik Garrison and Gabor Marth.
\newblock {Haplotype-based variant detection from short-read sequencing}, 2012.

\bibitem{dunn2023vcfdist}
Tim Dunn and Satish Narayanasamy.
\newblock {vcfdist: Accurately Benchmarking Phased Small Variant Calls in Human Genomes}.
\newblock {\em Nature Communications}, 2023.

\bibitem{giab_benchmark_hg002}
{Genome in a Bottle}.
\newblock {HG002 GRCh38 v4.1 Small Variant Draft Benchmark VCF}.
\newblock \url{https://ftp.ncbi.nlm.nih.gov/ReferenceSamples/giab/data/AshkenazimTrio/analysis/NIST_v4.1_SmallVariantDraftBenchmark_12182019/GRCh38/HG002_GRCh38_1_22_v4.1_draft_benchmark.vcf.gz}, 2019.

\bibitem{hwang2020hotchips}
Sangyun Hwang.
\newblock {Poster Presentation at Hot Chips 32: A Symposium on High Performance Chips}.
\newblock \url{https://www.hc32.hotchips.org/assets/program/posters/HC2020.SamsungFoundry.SangyunHwang.v03.pdf}, 2020.

\bibitem{locuza_nvidia_ada}
Locuza.
\newblock {NVIDIA's Ada Lineup \& Configurations}.
\newblock {\em Substack}, Sep 2023.
\newblock Blog post.

\bibitem{chao1992aligning}
Kun-Mao Chao, William~R Pearson, and Webb Miller.
\newblock {Aligning Two Sequences within a Specified Diagonal Band}.
\newblock {\em Bioinformatics}, 1992.

\bibitem{harris2007banded}
Brandon Harris, Arpith~C Jacob, Joseph~M Lancaster, Jeremy Buhler, and Roger~D Chamberlain.
\newblock A banded smith-waterman fpga accelerator for mercury blastp.
\newblock In {\em 2007 International Conference on Field Programmable Logic and Applications}, pages 765--769. IEEE, 2007.

\bibitem{ARM_AXI4Stream}
{Arm Ltd.}
\newblock Amba axi4-stream protocol specification.
\newblock \url{https://developer.arm.com/documentation/ihi0051}, 2013.
\newblock AMBA Specification IHI 0051.

\bibitem{PCIeGen3}
{PCI-SIG}.
\newblock Pci express base specification revision 3.0.
\newblock \url{https://pcisig.com/specifications}, 2010.
\newblock PCI Express Gen 3, 8.0 GT/s.

\bibitem{PCIeGen4}
{PCI-SIG}.
\newblock Pci express base specification revision 4.0.
\newblock \url{https://pcisig.com/specifications}, 2017.
\newblock PCI Express Gen 4, 16.0 GT/s.

\bibitem{Wagner2022}
Justin Wagner, Nathan~D. Olson, Lindsay Harris, Jennifer McDaniel, Haoyu Cheng, Arkarachai Fungtammasan, Yih-Chii Hwang, Richa Gupta, Aaron~M. Wenger, William~J. Rowell, Ziad~M. Khan, Jesse Farek, Yiming Zhu, Aishwarya Pisupati, Medhat Mahmoud, Chunlin Xiao, Byunggil Yoo, Sayed Mohammad~Ebrahim Sahraeian, Danny~E. Miller, David J{\'a}spez, Jos{\'e}~M. Lorenzo-Salazar, Adri{\'a}n Mu{\~{n}}oz-Barrera, Luis~A. Rubio-Rodr{\'i}guez, Carlos Flores, Giuseppe Narzisi, Uday~Shanker Evani, Wayne~E. Clarke, Joyce Lee, Christopher~E. Mason, Stephen~E. Lincoln, Karen~H. Miga, Mark T.~W. Ebbert, Alaina Shumate, Heng Li, Chen-Shan Chin, Justin~M. Zook, and Fritz~J. Sedlazeck.
\newblock {Curated Variation Benchmarks for Challenging Medically Relevant Autosomal Genes}.
\newblock {\em Nature Biotechnology}, 2022.

\bibitem{Barbitoff2022}
Yury~A. Barbitoff, Ruslan Abasov, Varvara~E. Tvorogova, Andrey~S. Glotov, and Alexander~V. Predeus.
\newblock {Systematic Benchmark of state-of-the-art Variant Calling Pipelines Identifies Major Factors Affecting Accuracy of Coding Sequence Variant Discovery}.
\newblock {\em BMC Genomics}, 2022.

\bibitem{Holtgrewe2010Mason}
Manuel Holtgrewe.
\newblock {Mason — A Read Simulator for Second Generation Sequencing Data}.
\newblock Technical report, Freie Universität Berlin, 2010.
\newblock Submitted version, deposited in FU Berlin repository.

\bibitem{Cheng2024}
Chu Cheng, Qingzhou Cheng, Wei Zhou, Yulong Chen, Wenbin Liu, Zhiling Zhang, Jingsi Ye, and Pengfeng Xiao.
\newblock {A correctable Decoding DNA Sequencing with High Accuracy and High Throughput}.
\newblock {\em Analytical Methods}, 2024.

\bibitem{Sun2024}
Jianfeng Sun, Martin Philpott, Danson Loi, Shuang Li, Pablo Monteagudo-Mesas, Gabriela Hoffman, Jonathan Robson, Neelam Mehta, Vicki Gamble, Tom~Brown Jr, Tom~Brown Sr, Stefan Canzar, Udo Oppermann, and Adam~P. Cribbs.
\newblock {Correcting PCR Amplification Errors in Unique Molecular Identifiers to Generate Accurate Numbers of Sequencing Molecules}.
\newblock {\em Nature Methods}, 2024.

\bibitem{10396343}
Zheqi Cao, Jinqiu Xu, and Jian Li.
\newblock {Minimap2 Parallelization Method based on Distributed Computing}.
\newblock In {\em ICHIH}, 2023.

\bibitem{koboldt2020best}
Daniel~C Koboldt.
\newblock {Best Practices for Variant Calling in Clinical Sequencing}.
\newblock {\em Genome medicine}, 2020.

\bibitem{meynert2014variant}
Alison~M Meynert, Morad Ansari, David~R FitzPatrick, and Martin~S Taylor.
\newblock {Variant Detection Sensitivity and Biases in Whole Genome and Exome Sequencing}.
\newblock {\em BMC bioinformatics}, 2014.

\bibitem{10177185}
Chung-Hsuan Yang, Yi-Chung Wu, Yen-Lung Chen, Chao-Hsi Lee, Jui-Hung Hung, and Chia-Hsiang Yang.
\newblock {An FM-Index Based High-Throughput Memory-Efficient FPGA Accelerator for Paired-End Short-Read Mapping}.
\newblock {\em IEEE Transactions on Biomedical Circuits and Systems}, 2023.

\bibitem{dunn2021squigglefilter}
Tim Dunn, Harisankar Sadasivan, Jack Wadden, Kush Goliya, Kuan-Yu Chen, David Blaauw, Reetuparna Das, and Satish Narayanasamy.
\newblock {Squigglefilter: An Accelerator for Portable Virus Detection}.
\newblock In {\em MICRO}, 2021.

\bibitem{HOUTGAST201854}
Ernst~Joachim Houtgast, Vlad-Mihai Sima, Koen Bertels, and Zaid Al-Ars.
\newblock {Hardware acceleration of BWA-MEM genomic short read mapping for longer read lengths}.
\newblock {\em Computational Biology and Chemistry}, 2018.

\bibitem{6972452}
Pei Liu, Ahmed Hemani, and Kolin Paul.
\newblock {A Many-Core Hardware Acceleration Platform for Short Read Mapping Problem using Distributed Memory Interface with 3D-stacked Architecture}.
\newblock In {\em SoC}, 2014.

\bibitem{Kim2018}
Jeremie~S. Kim, Damla~Senol Cali, Hongyi Xin, Donghyuk Lee, Saugata Ghose, Mohammed Alser, Hasan Hassan, Oguz Ergin, Can Alkan, and Onur Mutlu.
\newblock {GRIM-Filter: Fast Seed Location Filtering in DNA Read Mapping using Processing-In-Memory Technologies}.
\newblock {\em BMC Genomics}, 2018.

\bibitem{10568393}
Zhen Ju, Jingjing Zhang, Xuelei Li, Jintao Meng, and Yanjie Wei.
\newblock {SeedHit: A GPU Friendly Pre-Align Filtering Algorithm}.
\newblock {\em TCBB}, 2024.

\bibitem{bingol2021gatekeeper}
Z{\"u}lal Bing{\"o}l, Mohammed Alser, Onur Mutlu, Ozcan Ozturk, and Can Alkan.
\newblock {GateKeeper-GPU: Fast and Accurate Pre-Alignment Filtering in Short Read Mapping}.
\newblock {\em IPDPSW}, 2021.

\bibitem{Sadasivan2023-qx}
Harisankar Sadasivan, Milos Maric, Eric Dawson, Vishanth Iyer, Johnny Israeli, and Satish Narayanasamy.
\newblock {Accelerating Minimap2 for Accurate Long Read Alignment on {GPUs}}.
\newblock {\em J Biotechnol Biomed}, 2023.

\bibitem{Dong2024.03.23.586366}
Juechu Dong, Xueshen Liu, Harisankar Sadasivan, Sriranjani Sitaraman, and Satish Narayanasamy.
\newblock {mm2-gb: GPU Accelerated Minimap2 for Long Read DNA Mapping}.
\newblock {\em bioRxiv}, 2024.

\bibitem{marco2021fast}
Santiago Marco-Sola, Juan~Carlos Moure, Miquel Moreto, and Antonio Espinosa.
\newblock {Fast Gap-affine Pairwise Alignment Using the Wavefront Algorithm}.
\newblock {\em Bioinformatics}, 2021.

\bibitem{5161066}
Gregory~M. Striemer and Ali Akoglu.
\newblock {Sequence Alignment with GPU: Performance and Design Challenges}.
\newblock In {\em {IPDPS}}, 2009.

\bibitem{10.1145/3698587.3701366}
Juechu Dong, Xueshen Liu, Harisankar Sadasivan, Sriranjani Sitaraman, and Satish Narayanasamy.
\newblock {mm2-gb: GPU Accelerated Minimap2 for Long Read DNA Mapping}.
\newblock In {\em BCB}, 2024.

\bibitem{madhavan2014race}
Advait Madhavan, Timothy Sherwood, and Dmitri Strukov.
\newblock {Race Logic: A Hardware Acceleration for Dynamic Programming Algorithms}.
\newblock {\em ISCA}, 2014.

\bibitem{10.1145/3613424.3614306}
Max Doblas, Oscar Lostes-Cazorla, Quim Aguado-Puig, Nick Cebry, Pau Fontova-Must\'{e}, Christopher~Frances Batten, Santiago Marco-Sola, and Miquel Moret\'{o}.
\newblock {GMX: Instruction Set Extensions for Fast, Scalable, and Efficient Genome Sequence Alignment}.
\newblock In {\em MICRO}, 2023.

\bibitem{10609714}
Julian Pavon, Ivan~Vargas Valdivieso, Carlos Rojas, Cesar Hernandez, Mehmet Aslan, Roger Figueras, Yichao Yuan, Joël Lindegger, Mohammed Alser, Francesc Moll, Santiago Marco-Sola, Oguz Ergin, Nishil Talati, Onur Mutlu, Osman Unsal, Mateo Valero, and Adrian Cristal.
\newblock {QUETZAL: Vector Acceleration Framework for Modern Genome Sequence Analysis Algorithms}.
\newblock In {\em ISCA}, 2024.

\bibitem{10476438}
Sumit Walia, Cheng Ye, Arkid Bera, Dhruvi Lodhavia, and Yatish Turakhia.
\newblock {TALCO: Tiling Genome Sequence Alignment Using Convergence of Traceback Pointers}.
\newblock In {\em HPCA}, 2024.

\end{thebibliography}

\end{document}